\documentclass[fleqn,usenatbib]{mnras}

\usepackage{newtxtext,newtxmath}

\usepackage[utf8]{inputenc}
\usepackage[T1]{fontenc}
\usepackage{subcaption}
\usepackage{float}
\usepackage{placeins} 
\usepackage{soul}
\DeclareRobustCommand{\VAN}[3]{#2}
\let\VANthebibliography\thebibliography
\def\thebibliography{\DeclareRobustCommand{\VAN}[3]{##3}\VANthebibliography}

\usepackage{graphicx}	
\usepackage{amsmath}	

\newcommand{\Msun}{\mbox{$\rm M_{\odot}$}} 

\title[Origin of CO-rich hot subdwarfs]{On the origin of carbon- and oxygen-rich hot subdwarfs: an sdB merger scenario}

\author[B.Tang and X.Ji et al.]{
Bo Tang,$^{1,2}$\thanks{These authors contributed equally to this work.}
Xinyu Ji,$^{1,2}$\footnotemark[1]
Songyu Hu,$^{1,2}$
Zhongyang Liu,$^{1,2}$
and
Xianfei Zhang$^{1,2}$\thanks{Corresponding author: zxf@bnu.edu.cn}
\\
$^{1}$Institute for Frontier in Astronomy and Astrophysics, Beijing Normal University, Beijing, 102206, China\\
$^{2}$School of Physics and Astronomy, Beijing Normal University, Beijing, 100875, China
}

\date{Accepted XXX. Received YYY; in original form ZZZ}

\pubyear{\the\year{}}

\begin{document}
\label{firstpage}
\pagerange{\pageref{firstpage}--\pageref{lastpage}}
\maketitle

\begin{abstract}
Carbon- and oxygen-rich hot subdwarfs (CO-sdO stars) represent an extremely rare subclass of helium-rich hot subdwarfs. Recent observations of PG 1654+322, PG 1528+025, and UCAC4 108 reveal unusually large surface mass fractions of carbon and oxygen, which are difficult to explain within existing formation scenarios. We explore whether mergers involving sdB stars can provide a viable formation channel for these objects. Using the stellar evolution code MESA, we model the post-merger evolution of two merger channels: mergers between two sdB stars and mergers between a helium white dwarf and a sdB star. Carbon and oxygen synthesized during core helium burning in sdB stars may provide a natural chemical reservoir if efficient mixing occurs during the merger. The resulting evolutionary tracks are broadly consistent with the effective temperatures, surface gravities, and luminosities observed for the three stars, suggesting remnant masses of about 0.8–1.0 $\Msun$. Our calculations suggest that mergers involving sdB stars may represent a viable formation pathway for CO-rich hot subdwarfs.
\end{abstract}




\section{Introduction}
Hot subdwarfs are a class of compact, hot stars whose spectra are classified primarily as type B (sdB) or type O (sdO), with typical effective temperatures between 20,000 and 80,000~K and surface gravities $\log(g)\simeq4.5$--$6.5$. Canonical sdB stars typically have masses of about $0.5\,M_\odot$ and consist of a core-helium-burning star surrounded by a very thin hydrogen envelope ($M_{\rm env}<0.01\,M_\odot$; \citealt{Heber2009,Heber2016}). After core-helium exhaustion, these canonical sdB stars generally do not ascend the asymptotic giant branch. Instead, they evolve toward the white-dwarf cooling sequence and eventually become CO white dwarfs. The broader hot-subdwarf population, however, can include hotter, more luminous, and more massive objects with different evolutionary histories.

Because of their extremely high surface gravity and the effects of elemental diffusion, the surfaces of hot subdwarfs are predominantly composed of hydrogen. However, a subset of these stars exhibits helium-rich atmospheres, some of which are almost entirely composed of helium. Hot subdwarfs can therefore be broadly classified into hydrogen-rich (H-rich), intermediate-helium-rich (iHe), and helium-rich (He-rich) subtypes (e.g., as discussed in the works by \cite{Ahmad2003, Naslim2010, Luo2021, Luo2024,Lei2019, Lei2022}, and \cite{simon2021, simon2026}). For helium-rich hot subdwarfs, approximately half of them show significant carbon enrichment, with logarithm carbon mass fractions $\log(\beta_c)$ (where $\log(\beta_c)$ denotes the logarithm carbon mass fraction) corresponding to  $\log(\beta_c) \simeq$ -2.4 to -1.5 (see, e.g., \cite{Hirsch2009, Naslim2010}). The evolutionary origin of this enrichment remains poorly understood. In recent years, three helium-rich hot subdwarfs with extreme carbon enrichment have been discovered, e.g., PG~1654+322, PG~1528+025 and UCAC4~108 \citep{Werner2022,Werner2025}. In these stars, the carbon mass fraction reaches $25\%$ ($\log(\beta_c)\sim -0.6$), while oxygen is also enhanced, reaching $23\%$, far exceeding the levels seen in ordinary carbon-enriched hot subdwarfs. This poses a significant challenge for current hot subdwarf formation models.

Hot subdwarfs primarily form from low-mass stars that evolve off the red giant branch or have just experienced the helium flash at the tip of the red giant branch. Through various mechanisms, these stars lose most of their hydrogen envelope, leaving behind a helium-burning core. When helium ignites and burns stably in the core, the star becomes a hot subdwarf \citep{Heber2009,Heber2016}. A large fraction of stars are found in binary systems. During binary evolution, mass transfer through Roche-lobe overflow allows stars to lose or accrete material from their companions. For binaries containing a red giant, such mass transfer can lead to the formation of hot subdwarf systems, primarily through three evolutionary channels: stable Roche-lobe overflow, common-envelope ejection, and double white dwarf mergers \citep{Han2002,Han2003}. Observations indicate that many hot subdwarfs reside in binary systems, and their distributions in effective temperature, surface gravity, and orbital period are broadly consistent with theoretical predictions \citep{Geier2022}. Observationally, hot subdwarfs in wide binaries with long orbital periods are generally interpreted as products of the stable Roche-lobe overflow channel \citep{simon1998, Vos2020}. In contrast, close hot subdwarf binaries with orbital periods shorter than about 10 days are likely produced through the common-envelope ejection channel \citep{Geier2022, Schaffenroth2022, Schaffenroth2023}.

\begin{table}
    \centering
    \caption{Observed stellar parameters and surface abundances of the three known CO-sdO stars. Abundances are given as mass fractions.}
    \label{tab:parameter table of CO-sdOs}
    \begin{tabular}{lccc} 
        \hline
         & PG 1654+322$^{(1)}$ & PG 1528+025$^{(1)}$ &  UCAC4 108$^{(2)}$\\
        \hline
        $T_{\mathrm{eff}}$ (K) & 55000$\pm$3000 & 50000$\pm$3000  & 50000$\pm$3000\\
        $\log(g/\mathrm{cm\,s}^{-2})$ & 5.8$\pm$0.5 & 5.3$\pm$0.5 & 5.3$\pm$0.2\\
        $\log(L/L_{\odot})$ & 2.62$^{+0.23}_{-0.22}$ & 3.06$^{+0.43}_{-0.49}$  & 2.71$^{+0.18}_{-0.17}$\\
        He & 0.62$\pm$0.11 & 0.58$^{+0.17}_{-0.22}$ &  0.85$\pm$0.05 \\
        C  & 0.15$\pm$0.05 & 0.25$\pm$0.10 &  0.11$\pm$0.03\\
        O & 0.23$\pm$0.06 & 0.17$^{+0.12}_{-0.07}$  & 0.03$\pm$0.02\\
        \hline
    \end{tabular}
    \begin{flushleft}
    {\footnotesize \textbf{References.}(1)\cite{Werner2022},(2)\cite{Werner2025}.}
    \end{flushleft}
\end{table}

For helium-rich hot subdwarfs, \cite{Webbink1984} and \cite{Iben1990} proposed that the merger of two helium white dwarfs could reignite helium burning, producing a hot subdwarf. \cite{Saio2000} and \cite{Zhang2012} used detailed stellar evolution calculations to model the evolution and accretion-induced helium ignition resulting from the merger of two nearly pure helium white dwarfs. Because the merger involves two stars composed almost entirely of helium, the surface of the remnant is expected to be helium-rich. During merging, helium burning via the triple-alpha process may transport newly synthesized carbon to the surface by convection, leading to a degree of carbon enrichment. Stellar evolution and population synthesis studies further indicate that the distribution of carbon-rich hot subdwarfs is strongly correlated with stellar mass (\cite{Zhang2012}, \cite{Yu2021}). Observationally, helium-rich hot subdwarfs are predominantly single stars. Their effective temperatures, abundance patterns, and in some cases strong magnetic fields may all be linked to a merger origin. However, as mentioned earlier, for CO-rich sdO stars, the observed level of carbon enrichment is difficult to explain by the formation and subsequent surface transport in a double-helium white-dwarf merger scenario.

To explain the formation of CO-sdO stars, \cite{Miller2022} performed a more in-depth study of PG~1654+322 and PG~1528+025. They proposed a specific formation channel, in which CO-rich sdOs originate from the merger of a more massive He white dwarf with a less massive CO white dwarf. In this process, the material from the CO-WD is disrupted and accreted onto the He WD, igniting the helium shell and thereby forming a hot subdwarf with a carbon- and oxygen-rich envelope. This model successfully explains the observed characteristics of these stars and is reasonably consistent with evolutionary tracks. However, because the accreted material originates from a CO white dwarf, its carbon and oxygen abundances are significantly higher than those observed in PG~1654 +322 and PG~1528 +025. Nevertheless, if sufficiently strong diffusion effects are taken into account, the core helium burning continues, the radius of the sdB star increases until core helium burning ends, and the star enters the shell helium-burning phase, which is found in short-period binary systems, typically formed through common-envelope evolution. In some of these systems, the companion to the hot subdwarf is a helium white dwarf (He WD). The binary system discussed in \cite{Miller2022} falls into this category. However, unlike the scenario in \cite{Miller2022}, in most He WD + sdB systems, the sdB star is slightly more massive than the He WD. \cite{Justham2011} therefore proposed that, after further evolution into a double white dwarf (He WD + CO WD ) system, (CO WD formed from the evolution of the sdB, with a radius smaller than that of the He WD), the CO WD would accrete material from the He WD, forming a helium-rich hot subdwarf. However, since its core consists of carbon and oxygen, the resulting object would resemble a hot subdwarf in a late evolutionary stage after core helium burning has ceased. Because detailed abundance calculations were not performed in that work, it remains unclear whether carbon and oxygen would be significantly enriched in the hot subdwarf formed from such a merger.

In He WD + sdB systems, as helium continues to burn in the core during further evolution, the radius of the sdB star will increase until core helium burning ends and the star enters the shell helium burning phase (sdB). During this sdB phase, the star’s radius becomes significantly larger than that of the He WD. If the two stars are sufficiently close, unstable mass transfer may occur before the system evolves into a double white dwarf, with the He WD accreting material from the sdB star. Similar to the scenarios proposed by \cite{Justham2011} and \cite{Miller2022}, this process is expected to produce a helium-rich hot subdwarf. Owing to the unstable mass transfer and possible helium flashes during accretion, it is plausible that the newly synthesized carbon and oxygen from the sdB core could become thoroughly mixed in the outer envelope, potentially forming a CO-sdO star.

Another possible scenario involves the merger of two sdB stars during the sdB phase, which could also lead to the formation of a CO-sdO star. Theoretically, double sdB systems can form through common-envelope evolution \citep{Han2002,Han2003,Justham2011}, and such systems have been observed, such as PG 1544+488 \citep{Ahmad2003,sener2014}.

In this work, we investigate whether merger remnants involving sdB stars can explain the observed properties of recently discovered CO-rich hot subdwarfs. We consider two possible channels. The first is the merger of two sdB stars. The second is the merger of a sdB star with a He white dwarf. We investigate whether these scenarios can reproduce both the CO-rich surface composition and the observed locations of the three stars in the Kiel and HR diagrams. We do not attempt to uniquely determine the formation channel of any individual object. Our aim is to assess whether sdB material can naturally provide the required CO-rich chemical reservoir. We also ask if these merger channels are broadly consistent with the observed properties of the three CO-sdO stars. In this sense, our calculations primarily serve as a test of the physical consistency of the sdB merger scenario. They help clarify its viability as a possible formation pathway for CO-rich hot subdwarfs.

The paper is organized as follows. In Section 2, we describe the numerical methods and model assumptions in our calculations. In Section 3, we present the evolutionary results of the merger models. We also compare these with the observed properties of the three CO-sdO stars. In Section 4, we discuss the implications and limitations of the proposed scenarios and summarize our main conclusions.

\section{Methods}
\subsection{The merger channels }
\begin{figure}
    \centering
    \includegraphics[width=\columnwidth]{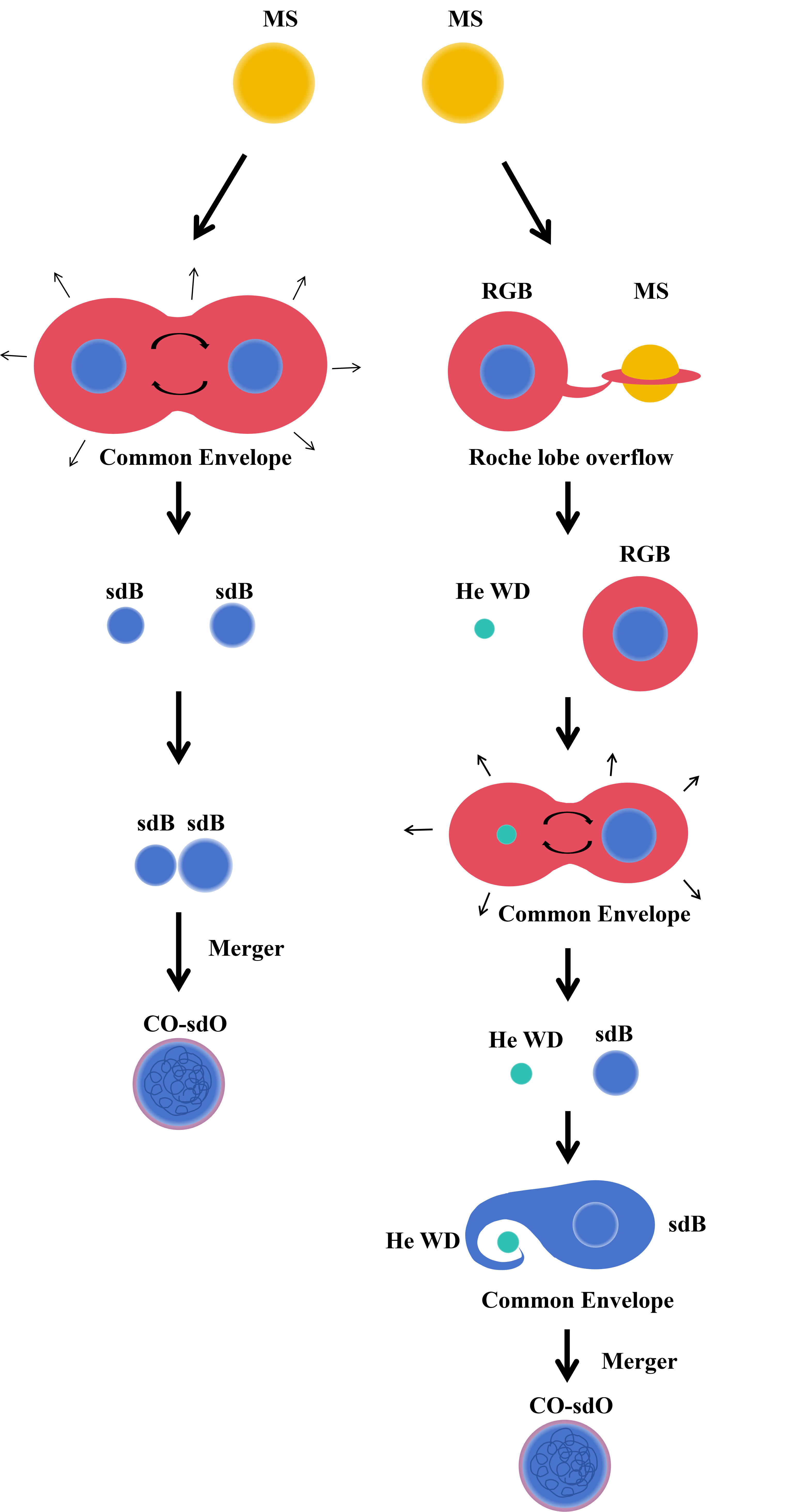}
    \caption{Proposed evolutionary scenarios for the formation of CO-sdO stars. The left panel illustrates the double sdB merger channel, while the right panel shows the He WD + sdB merger channel.}
    \label{fig:1}
\end{figure}

As illustrated in Fig.~\ref{fig:1}, we consider two possible formation channels for CO-sdO stars: the double sdB merger and the He WD + sdB merger.

(1) double sdB merger\\
Two main-sequence stars with comparable masses evolve into red giants and both expand to fill their Roche lobes, leading to the formation of a common envelope. The helium cores then orbit within the common envelope and gradually spiral inward, ejecting the surrounding hydrogen-rich material and leaving behind a binary composed of two helium cores. If both helium cores originate from sufficiently massive red giants, so that helium ignition is non-degenerate (see Fig. 1 of \cite{Han2002}), both stars can ignite helium and form a binary system of two hot subdwarfs with similar masses. During subsequent evolution, both stars may again fill their Roche lobes during the sdB phase, leading to a dynamical merger. If the merger leads to efficient mixing, helium burning can be re-established in the remnant, producing a massive hot subdwarf.

(2) He WD + sdB merger\\
Similar to the scenarios proposed by \cite{Justham2011} and \cite{Miller2022}, two main-sequence stars with different initial masses evolve as follows. The primary first becomes a red giant branch (RGB) star and, through Roche-lobe overflow, transfers its envelope to the companion, eventually forming a helium white dwarf (He WD). The companion subsequently evolves into a red giant, expands to fill its Roche lobe, undergoes unstable mass transfer, and forms a common envelope with the He WD. After the hydrogen envelope is ejected, a He WD + sdB binary is formed. During subsequent evolution, the radius of the sdB star increases significantly, triggering another phase of unstable mass transfer during which the sdB star is disrupted and accreted onto the He WD. This process produces a helium-rich envelope enriched in carbon and oxygen. Subsequent helium ignition then produces a CO-sdO star.

\subsection{The models }

To examine whether the two evolutionary channels illustrated in Fig.~\ref{fig:1} are consistent with the observed CO-sdO stars, we used the stellar evolution code MESA (Modules for Experiments in Stellar Astrophysics, version 24.08.01; \citealt{Paxton11,Paxton13,Paxton2015,Paxton2018}) to simulate post-merger evolution in both scenarios and compare the results with observational data. The main physical assumptions and parameter settings adopted in our calculations are as follows: parameters not explicitly specified are set to their MESA default values. For nuclear burning we adopt the \texttt{agb.net} reaction network, which includes 21 nuclides: $^{1}\mathrm{H}$, $^{2}\mathrm{H}$, $^{3}\mathrm{He}$, $^{4}\mathrm{He}$, $^{7}\mathrm{Li}$, $^{7}\mathrm{Be}$, $^{8}\mathrm{B}$, $^{12}\mathrm{C}$, $^{13}\mathrm{C}$, $^{13}\mathrm{N}$, $^{14}\mathrm{N}$, $^{15}\mathrm{N}$, $^{16}\mathrm{O}$, $^{17}\mathrm{O}$, $^{18}\mathrm{O}$, $^{19}\mathrm{F}$, $^{22}\mathrm{Ne}$, $^{23}\mathrm{Na}$, $^{24}\mathrm{Mg}$, $^{27}\mathrm{Al}$, $^{56}\mathrm{Fe}$. 

Convective instability is determined using the composition-dependent
stability criterion implemented in MESA. With
\texttt{use\_Ledoux\_criterion = .true.}, a radiative layer becomes
convectively unstable when
\begin{equation}
\nabla_{\rm rad}
-
\nabla_{\rm ad}
-
\frac{\phi}{\delta}\nabla_{\mu}
> 0,
\end{equation}
where $\nabla_{\rm rad}$ and $\nabla_{\rm ad}$ are the radiative and
adiabatic temperature gradients, respectively, and the last term
accounts for the composition gradient. Convection is treated using the
Cox MLT prescription with a mixing-length parameter of
$\alpha_{\rm MLT}=2.0$.

Semiconvective mixing is included using the Langer (1985)
prescription with $\alpha_{\rm sc}=0.1$. We adopt the MESA option
\texttt{`Langer\_85 mixing; gradT = gradr'}, for which the temperature
gradient in semiconvective regions is taken to be radiative,
$\nabla_T=\nabla_{\rm rad}$. Thermohaline mixing is treated using the
Kippenhahn et al. (1980) prescription with a thermohaline coefficient
of 1, and the thermal stratification in thermohaline regions is likewise
taken to be radiative, $\nabla_T=\nabla_{\rm rad}$.

Convective premixing is enabled in our MESA calculations to treat the evolution of convective boundaries. In addition, exponential overshooting is adopted. For convective cores, we use ($f=0.016$) and ($f_0=0.008$); for shell-burning regions, we use ($f=0.0174$) and ($f_0=0.0087$). Element diffusion is included throughout the calculations. The diffusion treatment follows the standard MESA implementation, with H, He, Li, C, N, O, and Fe used as representative diffusion species. These parameters, together with the complete MESA inlists, are provided in the Zenodo repository\footnotemark[1].

\textbf{\footnotetext[1]{These are available at \url{https://zenodo.org/records/20501687}.}}

The pre-merger models consist of hot subdwarfs and helium white dwarfs. During the merger, the high temperatures are expected to rapidly destroy the thin hydrogen-rich layers surrounding both the hot subdwarf and the helium white dwarf. To simplify the model setup and ensure numerical stability during the accretion calculations, all hot subdwarfs and helium white dwarfs are therefore assumed to have no hydrogen envelope.

For the sdB models, we consider masses of 0.4, 0.45, and 0.5 $\Msun$. These models are constructed from zero-age main-sequence stars with an initial mass of 2.5 $\Msun$ and metallicity $Z = 0.02$. This initial mass is chosen to produce a non-degenerate helium core, following Fig. 1 of \cite{Han2002}.
Once the helium core has grown to the desired mass during the red-giant phase, we use \texttt{relax\_mass} to remove nearly all of the hydrogen envelope, leaving a bare helium core. The stripped models are then evolved further to produce hot subdwarfs with masses of 0.4, 0.45, and 0.50 $\Msun$.

For the He WD models, we consider masses of 0.4, 0.45 $\Msun$. These models are constructed following an approach similar to that of \cite{Zhang2012}. We start from a zero-age main-sequence star with an initial mass of 1.2 $\Msun$ and metallicity  $Z = 0.02$, and evolve it until the helium core reaches 0.4 or 0.45 $\Msun$. As in the sdB models, the hydrogen envelope is then removed. Because the helium core is degenerate, central helium ignition does not occur, and the model cools directly to become a white dwarf. The evolution is terminated when the surface luminosity decreases to $\log(L/L_{\odot}) = -2$.

\begin{figure*}
    \centering
    \includegraphics[width=1\linewidth]{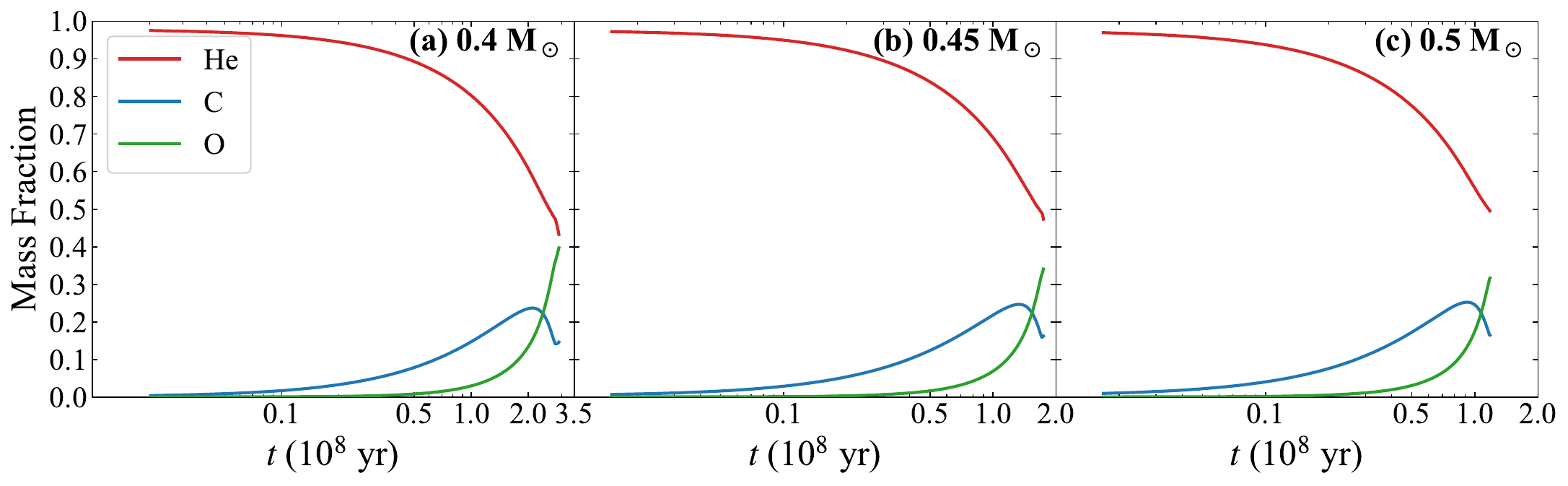}
    \caption{Evolution of the average mass fractions of He, C, and O for sdB stars with masses of 0.4, 0.45, and 0.5 $\Msun$ as a function of time.}
    \label{fig:2}
\end{figure*}

{For the double sdB merger models, we assume that the merger of two sdB stars is broadly analogous to the formation of blue stragglers. We therefore focus on the simplest limiting case in which the material from the two progenitors is fully mixed, producing a massive sdO star with a total mass equal to the sum of the progenitor masses. We use unevolved helium-star models only as initial hydrostatic templates for constructing the merger remnants. These initial templates are nearly pure helium objects and do not represent the final chemical composition of the merger products. We first rescale the template models to the desired remnant masses of 0.8, 0.85, 0.9, 0.95, and 1.0 $\Msun$ using \texttt{relax\_mass\_scale}. These masses correspond to merger events of $0.40+0.40, 0.40+0.45, 0.40+0.50, 0.45+0.45, 0.45+0.50,$ and $0.50+0.50$ $\Msun$, yielding six post-merger models. We then impose the observed abundance composition of three CO-sdO stars by replacing the abundances. The resulting models subsequently evolve into post-merger remnants, as detailed in Section 3.

For the He WD + sdB merger models, we adopt a scheme similar to that used by \cite{Zhang2012} and \cite{Yu2021}. Specifically, we simulate rapid accretion onto the helium white dwarf at prescribed high accretion rates of 10$^{-3}$ $\Msun$ yr$^{-1}$\citep{Yu2021}, chosen to reproduce the high-entropy envelope structure expected from merger calculations and smoothed particle hydrodynamics (SPH) simulations \citep{Dan2014}. The adopted accretion rates are not taken directly from SPH calculations; rather, they are chosen to reproduce a thermal structure similar to that of the merger remnant. This procedure produces a hot envelope structure, thereby yielding a post-merger remnant suitable for subsequent evolution calculations.

The composition of the accreted material is assumed to be identical to the fully mixed composition of the sdB donor. The final surface composition, therefore, depends on the mass and evolutionary stage of the sdB progenitor. Because the sdB star is assumed to be more massive than the He WD, we consider three representative merger combinations: 0.40 + 0.45, 0.40 + 0.50, and 0.45 + 0.50 $\Msun$.
These models are intended as one-dimensional approximations designed to test the physical consistency of the proposed merger channels, rather than fully self-consistent hydrodynamic simulations of the merger process.

\section{Results}
We first assess whether sdB material can provide He, C, and O abundances comparable to those observed in the three known CO-sdO stars. Having established the chemical feasibility of the proposed scenario, we then examine the double-sdB and He WD + sdB merger channels separately, focusing on their physical feasibility and on whether their post-merger evolutionary tracks can reproduce the observed stellar parameters.

\subsection{Average abundance of sdB}
In both the double sdB and He WD + sdB merger channels, the surface composition of the merger remnant (CO-sdO) is determined by material originating from the sdB star. The elemental abundances of sdB stars—which depend on both stellar mass and evolutionary stage—therefore play a key role in determining whether the merger remnants can reproduce the observed properties of CO-sdO stars.

\begin{figure}
    \includegraphics[width=\columnwidth]{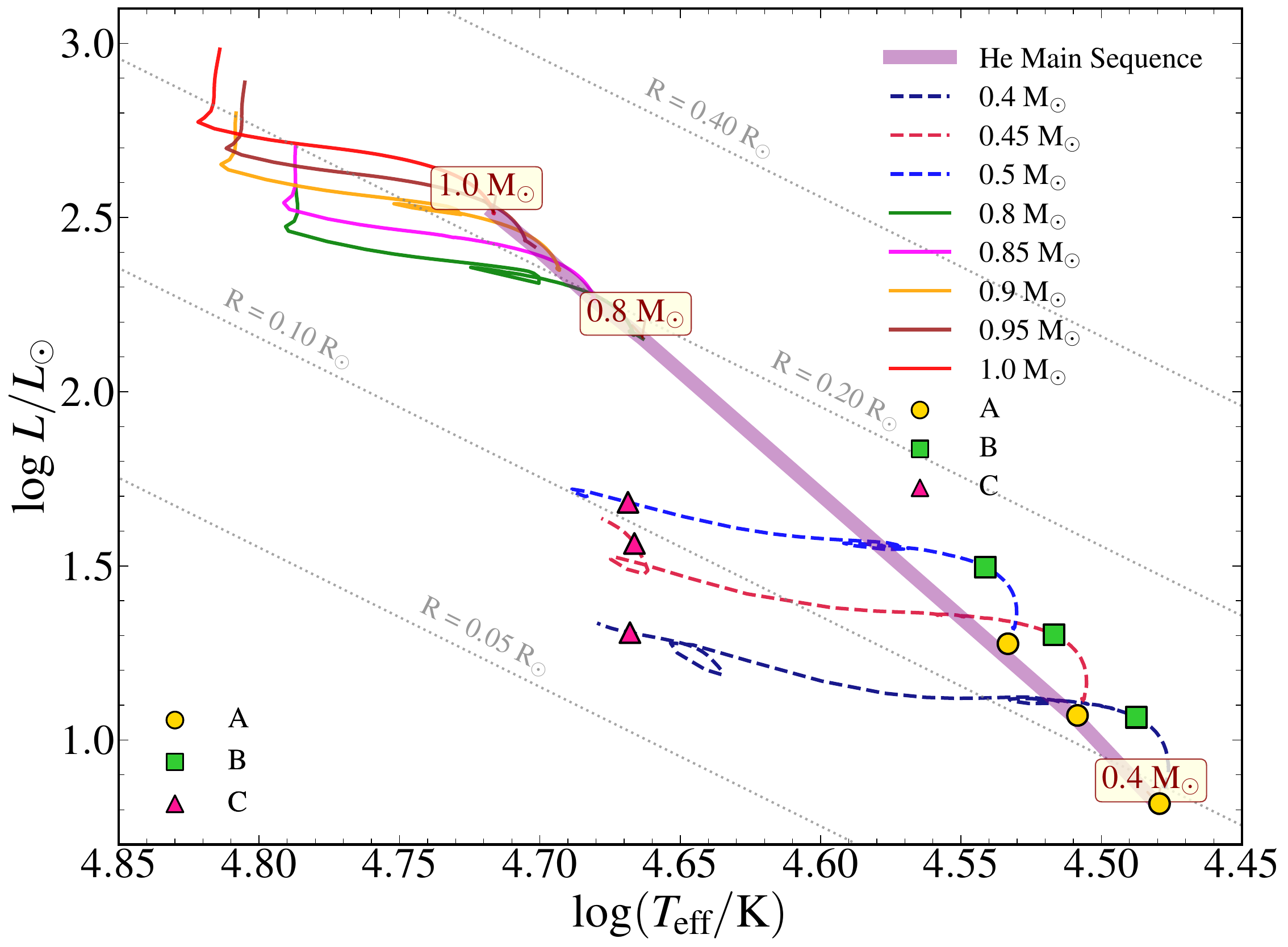}
    \caption{The Hertzsprung–Russell diagram illustrates the evolutionary tracks of sdB stars with varying masses. The purple band marks the helium main sequence, showing where core helium-burning sdB stars typically reside. Grey dashed lines indicate constant-radius tracks. Points A, B, and C represent key evolutionary stages: A is the start of the core helium-burning, B is the maximum radius, and C is the final stage shown.}
    \label{fig:3}
\end{figure}

\begin{figure}
    \includegraphics[width=\columnwidth]{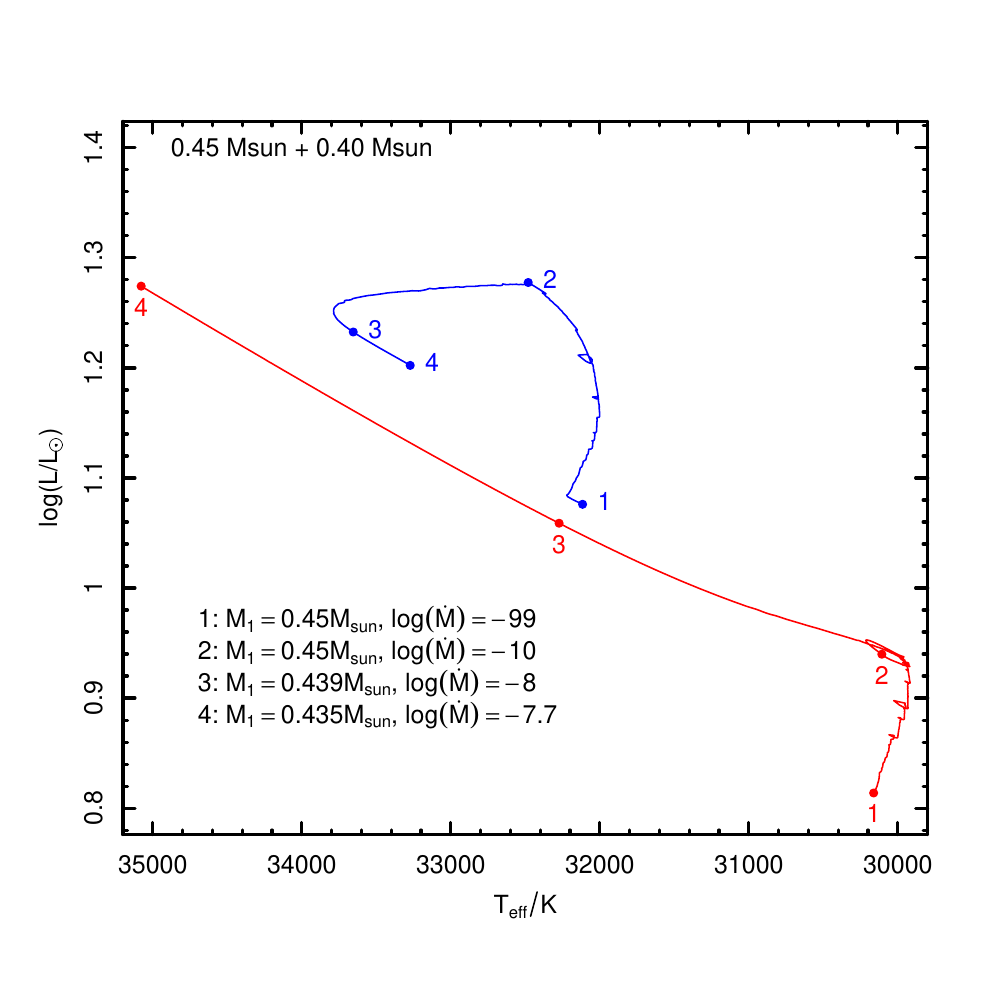}
\caption{
HR-diagram evolution of a representative compact double-sdB system with initial component masses of
\(0.45+0.40\,M_\odot\) and an initial separation of
\(0.60\,R_\odot\).
The blue and red curves show the evolutionary tracks of the primary and secondary, respectively.
Four characteristic stages are marked along the evolution.
}
\label{fig:4}
\end{figure}

\begin{figure}
    \includegraphics[width=\columnwidth]{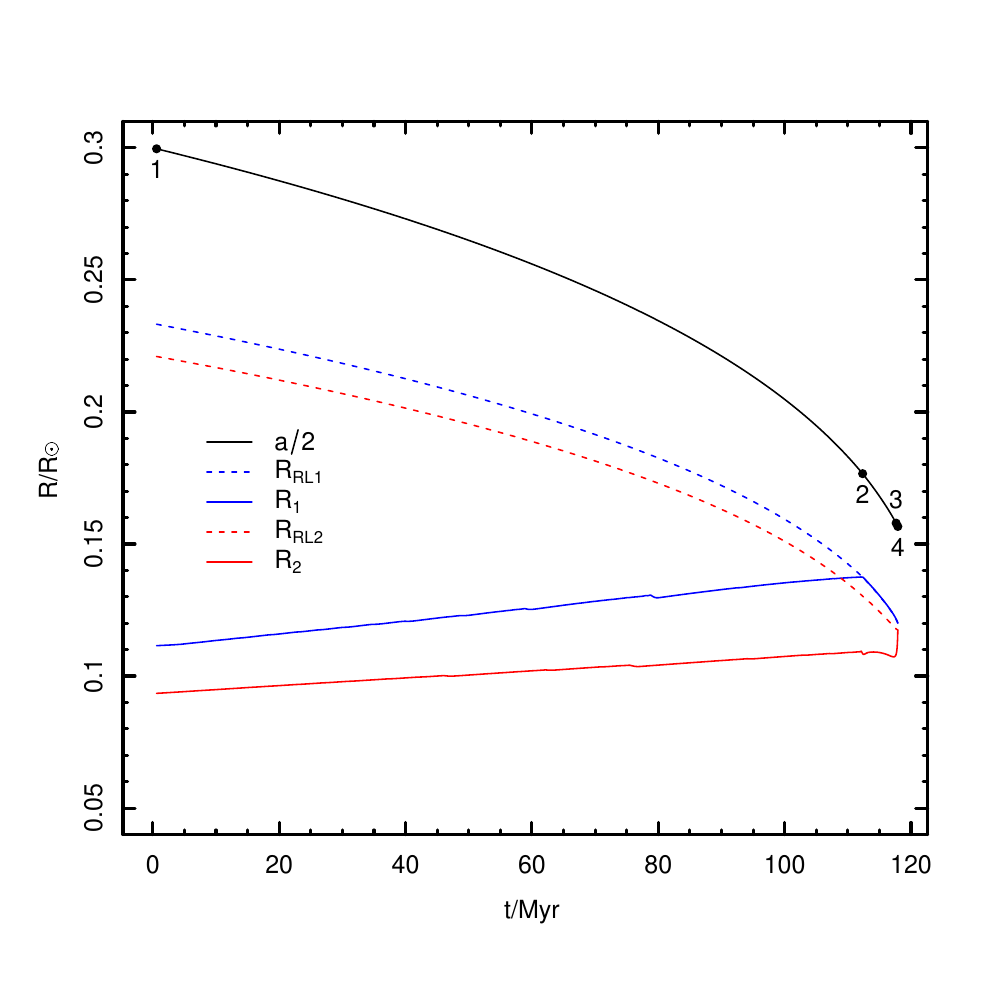}
\caption{Evolution of the orbital separation, stellar radii, and Roche-lobe radii for the same \(0.45+0.40\,M_\odot\) double-sdB system shown in Fig.~\ref{fig:4}. For clarity, the black curve denotes \(a/2\), rather than the full orbital separation \(a\). The blue and red solid curves show the radii of the primary and secondary, \(R_1\) and \(R_2\), respectively, while the corresponding dashed curves show their Roche-lobe radii, \(R_{\rm RL,1}\) and \(R_{\rm RL,2}\).
}
\label{fig:5}
\end{figure}

\begin{figure*}
    \centering
    \includegraphics[width=0.8\textwidth]{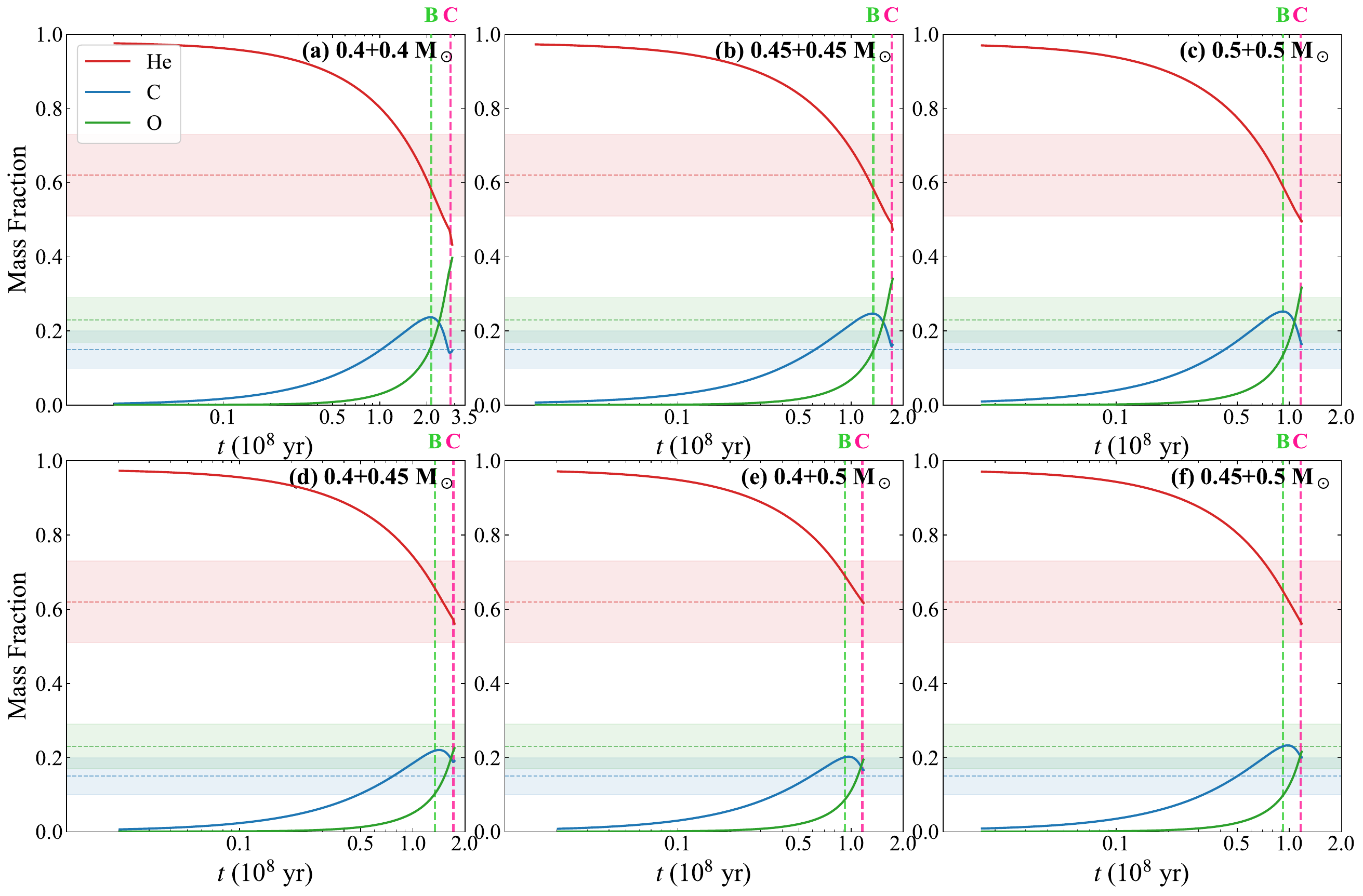}
    \vspace{0.2em}
    \caption{Panels (a)–(f) show how abundances of helium (He), carbon (C), and oxygen (O) evolve. Each panel presents results for one element in the merger scenario of PG 1654+322. Abundances assume fully mixed merger remnants, using the mass-weighted average of the two progenitors' initial compositions. Horizontal dashed lines mark the observed abundances for each element, and shaded regions indicate the observational uncertainties.}
    \label{fig:6}
\end{figure*}

\begin{figure*}
    \centering
    \includegraphics[width=0.8\textwidth]{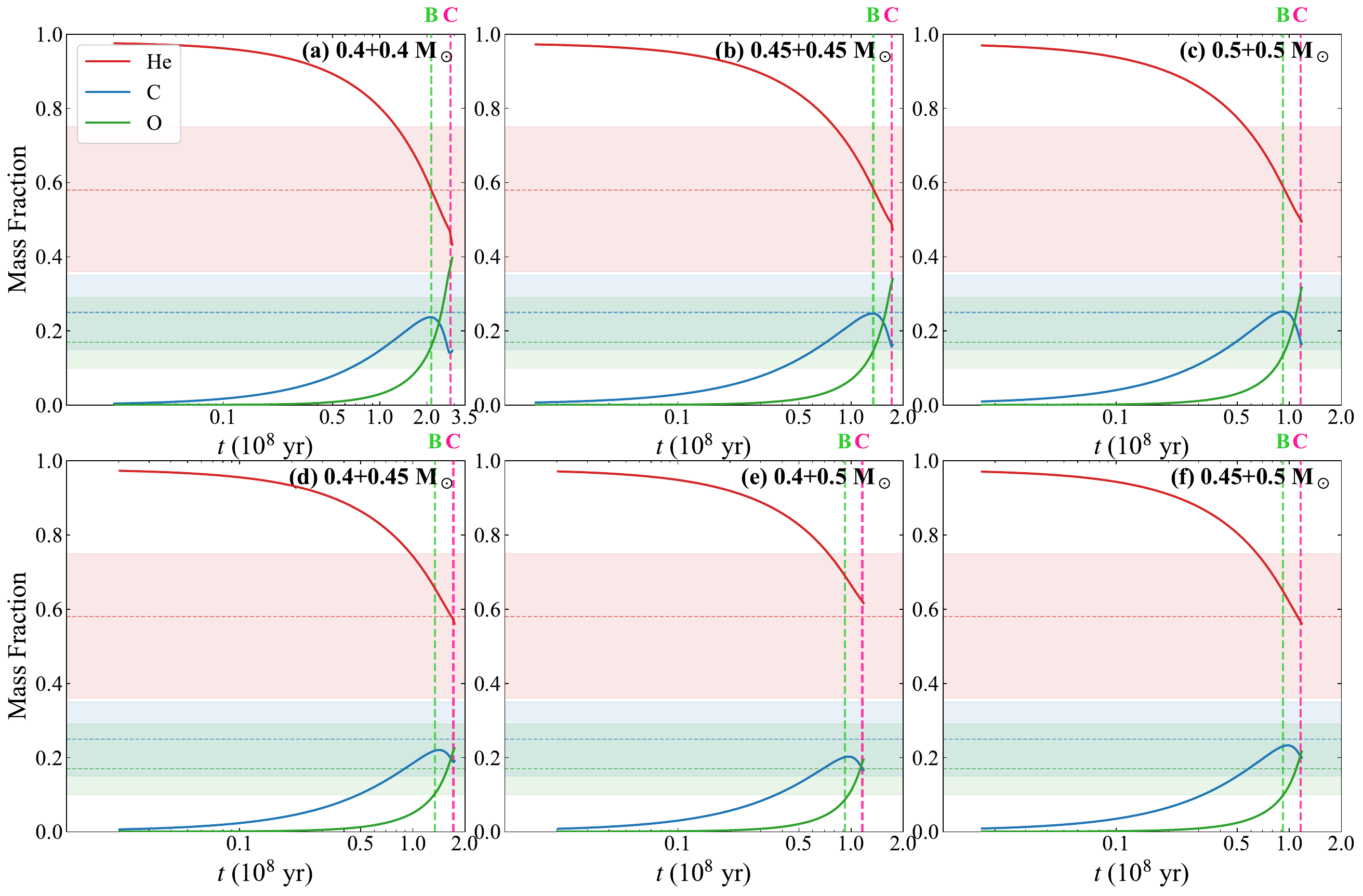}
    \vspace{0.2em}
    \caption{Same as Fig.~\ref{fig:6}, but for PG 1528+025.}
    \label{fig:7}
\end{figure*}

\begin{figure*}
    \centering
    \includegraphics[width=0.8\textwidth]{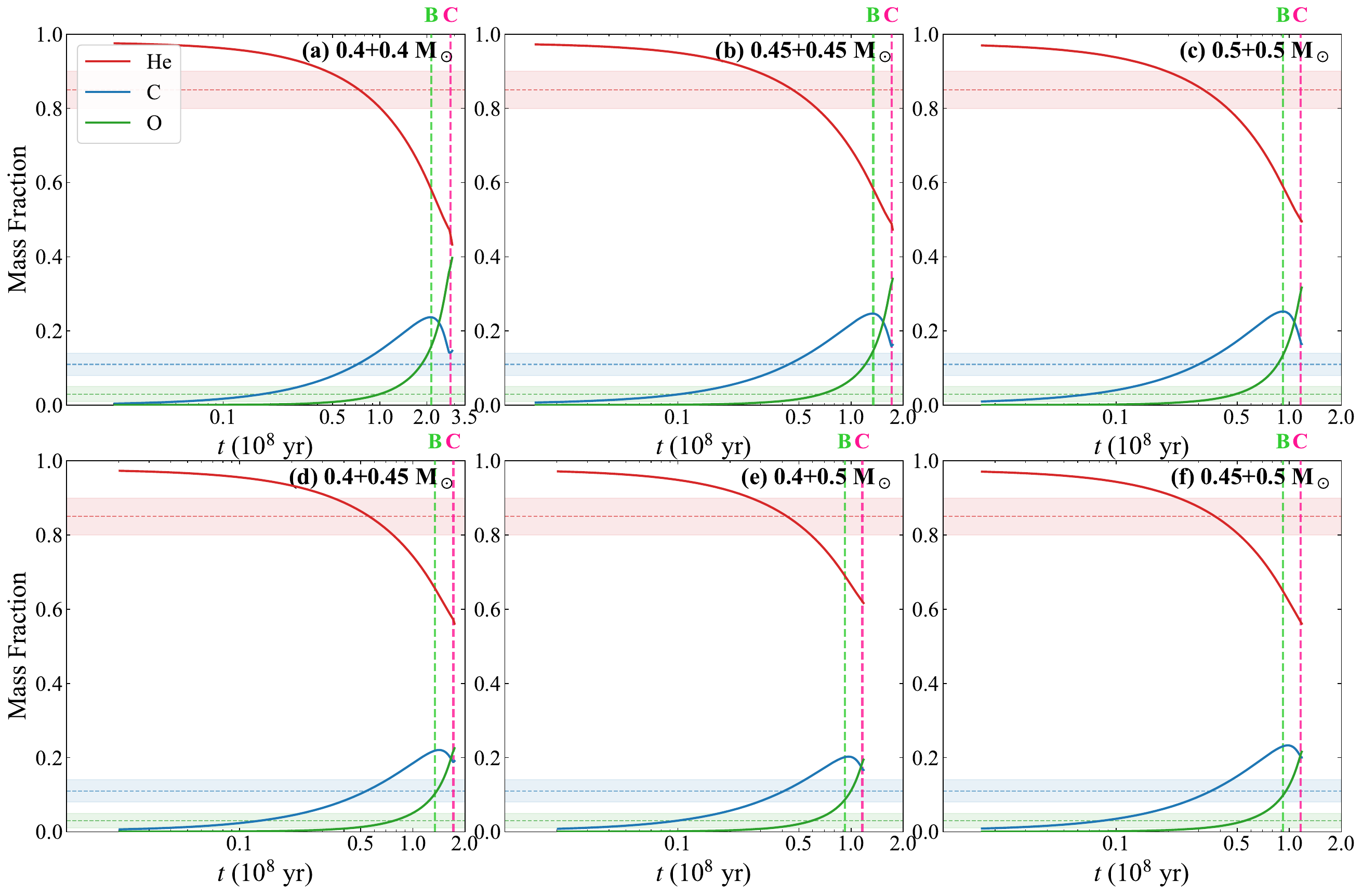}
    \vspace{0.2em}
    \caption{Same as Fig.~\ref{fig:6}, but for UCAC4 108.}
    \label{fig:8}
\end{figure*}

\begin{figure}
    \centering
    \includegraphics[width=1\linewidth]{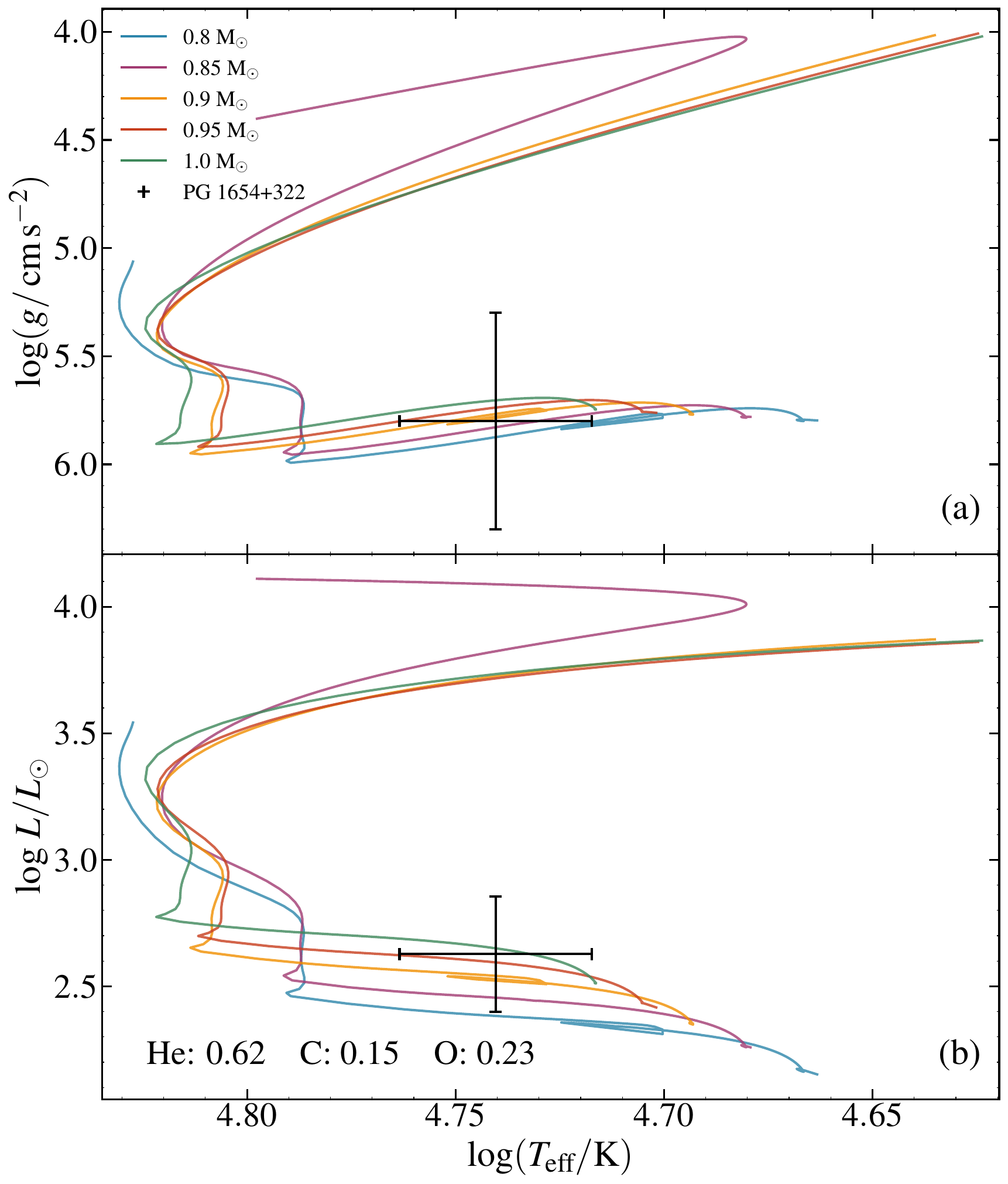}
    \caption{Evolutionary tracks for PG 1654+322 merger remnants in the double sdB merger scenario. Panel (a) shows the Kiel diagram, and panel (b) shows the HR diagram. Black error bars indicate the observed ranges of the stellar parameters.}
    \label{fig:9}
\end{figure}

\begin{figure}
    \centering
    \includegraphics[width=1\linewidth]{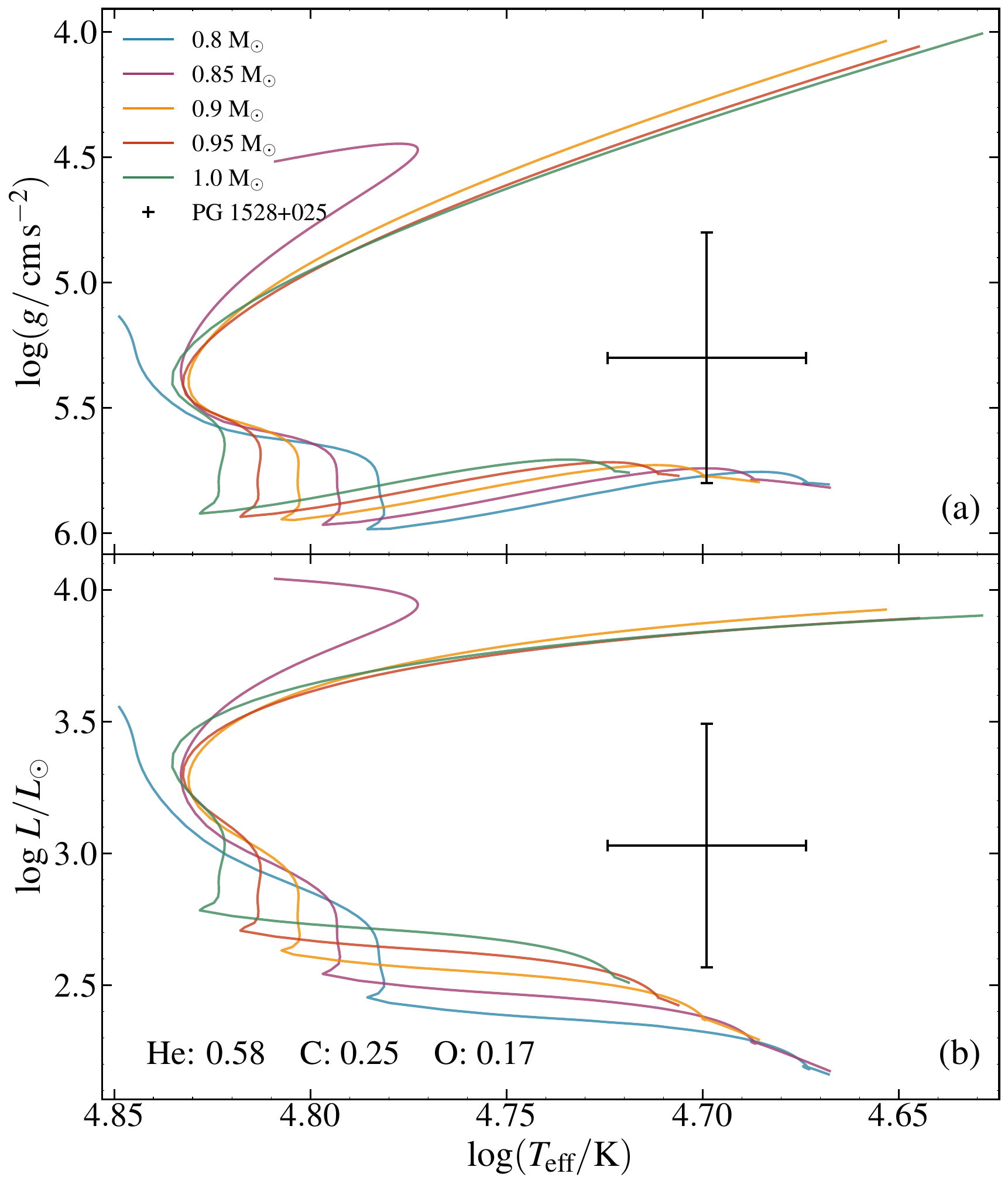}
    \caption{Same as Fig.~\ref{fig:9}, but for PG 1528+025.}
    \label{fig:10}
\end{figure}

\begin{figure}
    \centering
    \includegraphics[width=1\linewidth]{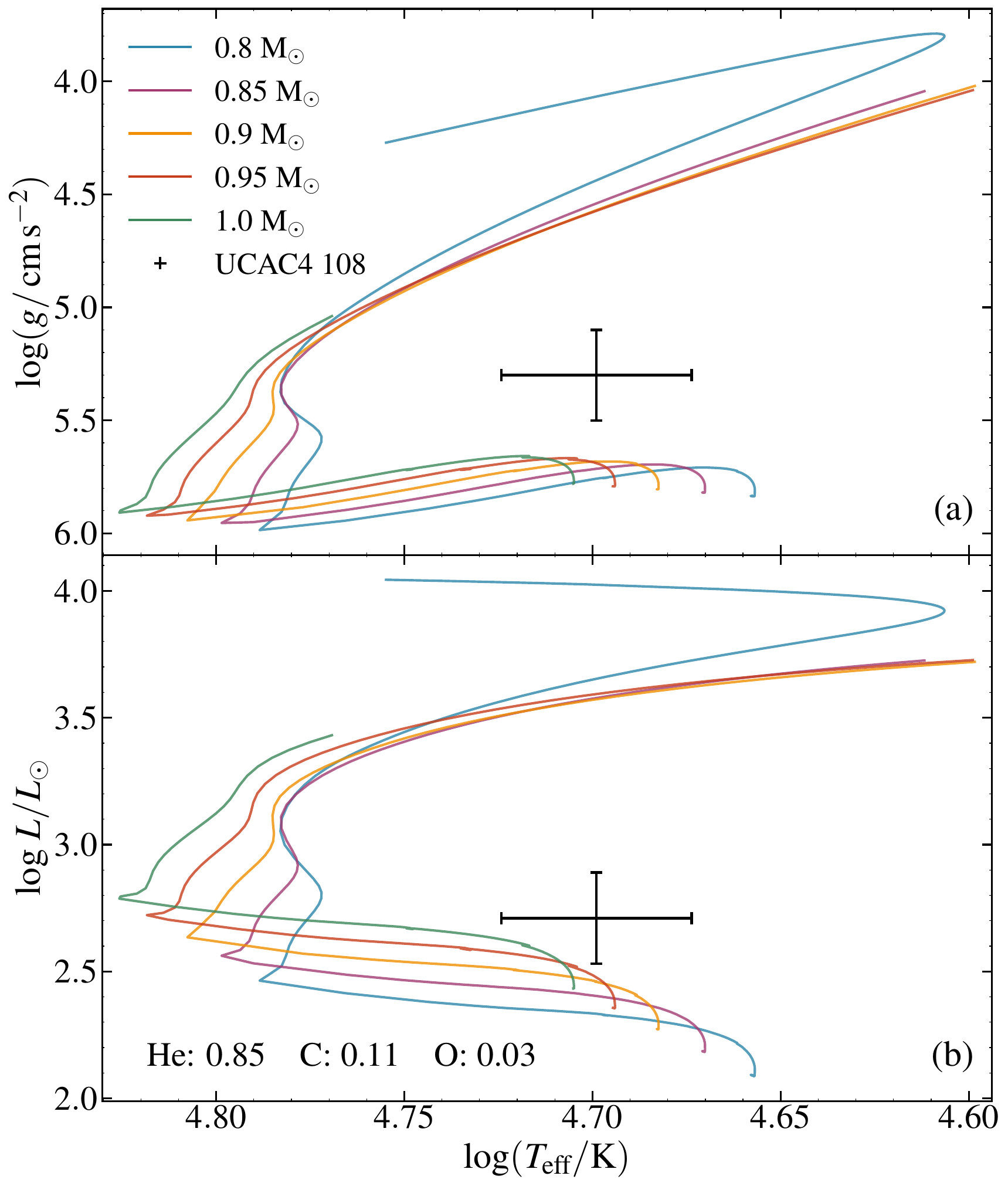}
    \caption{Same as Fig.~\ref{fig:9}, but for UCAC4 108.}
    \label{fig:11}
\end{figure}

As shown in Fig.~\ref{fig:2}, we calculate the average elemental abundances of sdB stars with masses of 0.4, 0.45, and 0.5 $\Msun$ as a function of time, from the helium zero-age main sequence (HeZAMS) to the end of core helium burning. These abundances represent the composition of the material that may participate in the merger. The figure shows that, as evolution proceeds, the abundances of carbon and oxygen increase steadily, leading to a corresponding rise in their average mass fractions. Conversely, the helium abundance gradually decreases with time. As a result, carbon and oxygen become significantly enhanced by about $10^7$ years after the onset of core helium burning. By $10^8$ years, the mass fractions of carbon and oxygen reach values of about 0.2, comparable to those observed in CO-sdO stars (Table~\ref{tab:parameter table of CO-sdOs}).

Another important factor affecting binary interaction is the evolution of the sdB stellar radius. As shown in Fig.~\ref{fig:3}, the radius of the hot subdwarf increases from point A to point B and then decreases towards point C. Therefore, if a merger occurs, mass transfer is most likely to take place during the A–B phase.

\subsection{sdB+sdB merger}

In the double sdB merger scenario, two hot subdwarfs—compact core-helium-burning stars that have lost most of their hydrogen envelopes—are assumed to form through common-envelope ejection. The two stars then evolve approximately synchronously. During the sdB phase, the stellar radius increases from point A to point B, as shown in Fig.~\ref{fig:3}. If the binary separation is sufficiently small, both stars may fill their Roche lobes during this expansion. This may lead to unstable mass transfer and eventually to a dynamical merger. The merger remnant is expected to form a more massive hot subdwarf, with a total mass approximately equal to the sum of the progenitor masses.

\subsubsection{Birth rate}

A key uncertainty in the proposed double-sdB merger channel is whether such compact double-sdB systems can be formed at all. In the standard picture, double-sdB systems may form through a double-core common-envelope event, in which two nearly coeval RGB stars enter a common envelope and eject their hydrogen-rich envelopes, leaving behind two non-degenerate helium cores that subsequently ignite helium. This evolutionary channel requires a very restricted range of initial masses and orbital separations, because both stars must develop helium cores and interact at nearly the same evolutionary stage. This requirement is particularly restrictive for non-degenerate helium cores, whose rapid post-main-sequence evolution requires the two progenitors to have nearly identical initial masses. Moreover, the detailed physics of double-core common-envelope evolution remains highly uncertain.

Based on the binary population-synthesis results of
\cite{Han2002,Han2003}, \citet{Justham2011} estimated the formation
rate of double-core hot-subdwarf systems by combining the
$\sim50\%$ close-binary fraction of sdB stars, the $\sim7\%$
fraction of close sdB binaries produced through the first
common-envelope channel with non-degenerate helium ignition, and
the $\sim1\%$ fraction of systems in this channel satisfying the
required nearly equal initial masses. This gives
$0.5\times0.07\times0.01\simeq3.5\times10^{-4}$,
corresponding to roughly one double-core system formed per
$3\times10^3$ normal sdB stars. This estimate should be regarded only as an order-of-magnitude estimate for the formation of double-core hot-subdwarf systems, rather than for double-sdB mergers themselves. Only a subset of these systems will have post-common-envelope separations sufficiently small to reach Roche-lobe contact during the core-helium-burning phase, so the fraction that ultimately merges is expected to be lower.

Observationally, PG~1544+488 \citep{Ahmad2003,sener2014} provides a direct proof of principle that close binaries containing two hot subdwarfs can form. It is particularly relevant to the present scenario because both components are helium-rich sdB stars with very similar properties. Its existence, however, does not by itself constrain how frequently such systems are produced.

\subsubsection{Stability of mass transfer}

The stability of mass transfer in compact double-sdB systems requires some caution. Previous studies suggest that mass transfer from radiative-envelope or helium-star donors is not necessarily dynamically unstable for near-unity mass ratios. \cite{Temmink2023} studied the stability of mass transfer from post-main-sequence donor stars and found that stable mass transfer is possible over a wider parameter space than commonly assumed in rapid binary population-synthesis calculations.
Using their convention, \(q=M_{\rm accretor}/M_{\rm donor}\), the quasi-adiabatic critical mass ratio for Hertzsprung-gap donors is of order \(q_{\rm qad}\simeq0.25\), corresponding to a donor-to-accretor mass ratio of order several. Similarly, the adiabatic mass-loss models for low- and intermediate-mass helium stars by \cite{zhang2024} give \(q_{\rm crit}=M_{\rm He}/M_{\rm accretor}\simeq1.0\)--\(2.6\) during the He-MS phase, while \(q_{\rm crit}\) increases rapidly after the early He-HG phase. These results imply that a near-equal-mass double-sdB binary is not expected to undergo dynamically unstable mass transfer solely because of its initial mass ratio.

However, these stability criteria mainly describe the response of the mass-losing donor on a dynamical or quasi-adiabatic timescale, and do not follow the detailed structural response of the accreting star. This limitation is particularly important for compact double-sdB binaries. For a circular binary, the characteristic gravitational-wave coalescence timescale can be written as
\begin{equation}
t_{\rm GW} =
\frac{5c^{5}a^{4}}
{256G^{3}M_{1}M_{2}(M_{1}+M_{2})},
\end{equation}
where $M_{1}$ and $M_{2}$ are the component masses and $a$ is the
orbital separation \citep{Postnov2014,Maoz2018}. For a representative
double-sdB binary with $M_{1}=0.45\,M_{\odot}, M_{2}=0.40\,M_{\odot}$ and
$a=0.60\,R_{\odot}$, this gives
$t_{\rm GW}\simeq 1.27\times10^{8}\,{\rm yr}$.
This timescale is comparable to the core-helium-burning lifetime of
an sdB star, indicating that gravitational-wave radiation can substantially reduce the orbital separation during the sdB phase.

Once Roche-lobe overflow begins, whether the accreting sdB star expands in response to mass accretion becomes important. A rapid increase in the mass-transfer rate may drive such systems to merge or undergo a common-envelope-like coalescence. Therefore, a self-consistent binary calculation including both the donor and the accretor is essential for assessing the final fate of compact double-sdB systems.

For these calculations, we used the binary module of MESA to evolve the two stellar components simultaneously. Both stars are treated as active stellar models, so that the radius evolution of the donor and the structural response of the accretor to mass accretion are followed self-consistently within the one-dimensional binary calculation. The Roche-lobe radii are evaluated during
the evolution, and Roche-lobe overflow is initiated when the stellar radius exceeds the corresponding Roche-lobe radius.

In the present exploratory calculations, the orbital angular-momentum loss is assumed to be driven only by gravitational-wave radiation. We do not include additional binary-interaction processes such as magnetic braking, tidal dissipation, spin-orbit coupling, or possible magnetic effects associated with the hot-subdwarf components. This simplification is appropriate for our
limited purpose of testing whether compact double-sdB systems can reach Roche-lobe contact during the core-He-burning lifetime, but it should not be interpreted as a complete treatment of the binary interaction. Additional angular-momentum loss or tidal effects could change the exact time of contact and the subsequent mass-transfer rate. Therefore, the binary calculations presented here should be regarded as representative feasibility tests rather than a full population-level prediction of the stability boundary.

To examine whether compact double-sdB systems can interact during the core-He-burning phase and to assess their subsequent fate, we performed detailed binary-evolution calculations for six representative systems with component masses of
$0.4+0.4$, $0.45+0.4$, $0.45+0.45$, $0.5+0.4$, $0.5+0.45$, and $0.5+0.5\,M_\odot$. In our calculations, systems with initial separations smaller than about $0.65\,R_\odot$ are driven towards Roche-lobe contact by gravitational-wave radiation on a timescale of the same order as, or shorter than, the core-He-burning lifetime.
A representative example is shown in Figs.~\ref{fig:4} and \ref{fig:5}, for a double-sdB binary with initial component masses of $0.45+0.40\,M_\odot$ and an initial separation of $0.60\,R_\odot$. As shown in Fig.~\ref{fig:4}, both components remain in the hot-subdwarf/core-He-burning region when Roche-lobe overflow begins, confirming that the interaction occurs during the double-sdB phase rather than after core-He exhaustion.

The corresponding evolution of the orbital separation, stellar radii, and Roche-lobe radii is shown in Fig.~\ref{fig:5}. For clarity, the plotted orbital-separation curve corresponds to $a/2$, rather than the full separation $a$. Gravitational-wave radiation continuously extracts orbital angular momentum, causing the binary separation and the Roche-lobe radii of both components to decrease with time.
Meanwhile, the radius of the primary gradually increases during its sdB evolution and reaches its Roche lobe at stage~2, initiating mass transfer. After the onset of Roche-lobe overflow, the accreting secondary expands in response to accretion. By stage~4, the secondary also fills its Roche lobe, and the system enters a double-contact configuration. At the same time, the mass-transfer rate has increased to $\log_{10}(|\dot{M}_1|/M_\odot\,{\rm yr}^{-1})\simeq -7.7$. This behavior indicates that the binary does not maintain a long-lived stable semi-detached mass-transfer phase. Thus, although near-equal-mass double-sdB systems may not be dynamically unstable solely according to the usual donor-response mass-transfer criteria, our detailed calculations show that gravitational-wave-driven orbital shrinkage, donor expansion, and the radial response of the accretor can jointly drive the system into double contact, making a subsequent merger or common-envelope-like coalescence the most probable outcome.

\subsubsection{Abundance and evolutionary tracks}

In principle, the surface abundances of the merger remnant depend on several factors, including the progenitors' masses and the evolutionary stage at which the merger occurs. Different combinations of these parameters would produce a large number of possible abundance evolution tracks. A fully predictive calculation of the surface abundances would therefore require detailed modelling of the merger process. However, the merger of two hot subdwarfs is a highly dynamical three-dimensional process involving complex hydrodynamics and mixing, which cannot currently be modelled self-consistently within one-dimensional stellar evolution calculations.

To obtain a first-order estimate, we therefore adopt a simplified approach in which the merger remnant is assumed to be fully mixed. Its composition is taken to be the mass-weighted average of the two progenitor sdB stars. We compute these mixed abundances for several representative binary mass combinations (0.40 + 0.40, 0.40 + 0.45, 0.40 + 0.50, 0.45 + 0.45, 0.45 + 0.50, and 0.50 + 0.50 $\Msun$).

Figures~\ref{fig:6}--\ref{fig:8} show that the observed abundance patterns of the three CO-sdO stars can be reproduced by several merger models, although the quality of the agreement depends on both the remnant mass and the evolutionary stage. PG 1654+322 is the easiest object to reproduce, with several models matching the observed abundances, typically at or near stage B. PG 1528+025 is more restrictive, requiring higher carbon abundances. We note that UCAC4 108 has a high helium abundance and a relatively low oxygen abundance, compared with PG 1654+322 and PG 1528+025, making its surface composition somewhat reminiscent of hydrogen-deficient, carbon-rich objects, including R CrB. This similarity may indicate that UCAC4 108 experienced a somewhat different mixing history or progenitor composition, although a detailed comparison with R CrB stars is beyond the scope of this work.

By comparing these results, we found that near point B in Fig.~\ref{fig:3}, the mixed He/C/O abundances for several mass combinations fall within the observational ranges of the three objects. However, performing post-merger evolutionary calculations for every possible combination of abundances would require an extremely large number of models. Given that only three CO-sdO stars are currently known, it is difficult to uniquely constrain the exact merger parameters.

Instead of predicting the post-merger surface abundances directly, we adopt an inverse approach: we use the observed abundances of the three CO-sdO stars (carbon- and oxygen-rich subdwarf O-type stars) to represent the compositions of merger remnants and compute their subsequent evolutionary tracks. This allows us to test whether such compositions can reproduce the observed stellar parameters, including effective temperature, surface gravity, and luminosity.

Our calculations should therefore be regarded as a consistency test of the merger scenario rather than a fully predictive model of the surface abundances.

The resulting evolutionary tracks are shown in Figs.~\ref{fig:9}--\ref{fig:11}. In the Kiel diagram, all three stars fall reasonably close to the theoretical tracks for merger remnants with masses between approximately 0.8 and 1.0 $\Msun$. In the HR diagram, PG 1654+322 is consistent with a stable core helium-burning phase. PG 1528+025 and UCAC4 108 appear somewhat more luminous than expected for central helium burning. This suggests they may already have entered a helium-shell-burning phase.

\subsection{He WD + sdB}

\subsubsection{Evolutionary tracks}
In the He WD + sdB merger scenario, the helium white dwarf is assumed to accrete material from the sdB star. Accretion of helium-rich material onto the He WD can trigger helium-shell ignition. Detailed stellar evolution studies by \cite{Zhang2012} and \cite{Yu2021} show that such accretion may produce a sequence of inward-propagating helium flashes, eventually leading to central helium ignition and the formation of a hot-subdwarf-like object.

The merger between a He WD and a sdB star is inherently a three-dimensional hydrodynamic process. At present, no dedicated hydrodynamic simulations exist for this type of merger. The accretion phase, therefore, cannot be modelled self-consistently within a one-dimensional stellar evolution framework. Instead, we approximate the merger by imposing a constant mass-accretion rate onto the He WD.

Following the approach adopted in previous studies (e.g., \cite{Zhang2012}; \cite{Yu2021}), we use accretion rates chosen to reproduce a high-temperature envelope structure similar to that predicted by hydrodynamic merger simulations. These rates are not intended to represent the exact physical accretion rate during the merger. Their main role is to reproduce a plausible thermal structure in the merger remnant. 

The chemical composition of the accreted material is assumed to be identical to that of the sdB progenitor. In principle, additional nucleosynthesis may occur during the merger and subsequent accretion process. For example, mergers of double helium white dwarfs can produce moderate carbon enrichment during helium flashes. However, the three observed CO-sdO stars show extremely large carbon and oxygen abundances, significantly exceeding the enrichment typically expected from such processes. We therefore assume that the dominant source of carbon and oxygen originates from nucleosynthesis in the core of the sdB star prior to the merger, while additional enrichment during the merger itself is neglected as a first-order approximation.

As discussed earlier, the elemental composition of sdB stars varies with both stellar mass and evolutionary stage. Different merger times and progenitor masses would therefore produce a wide range of possible abundance patterns. Given that only three CO-sdO stars are currently known, it is difficult to uniquely constrain the merger parameters using abundance modelling alone.

\begin{figure}
    \centering
    \includegraphics[width=1\linewidth]{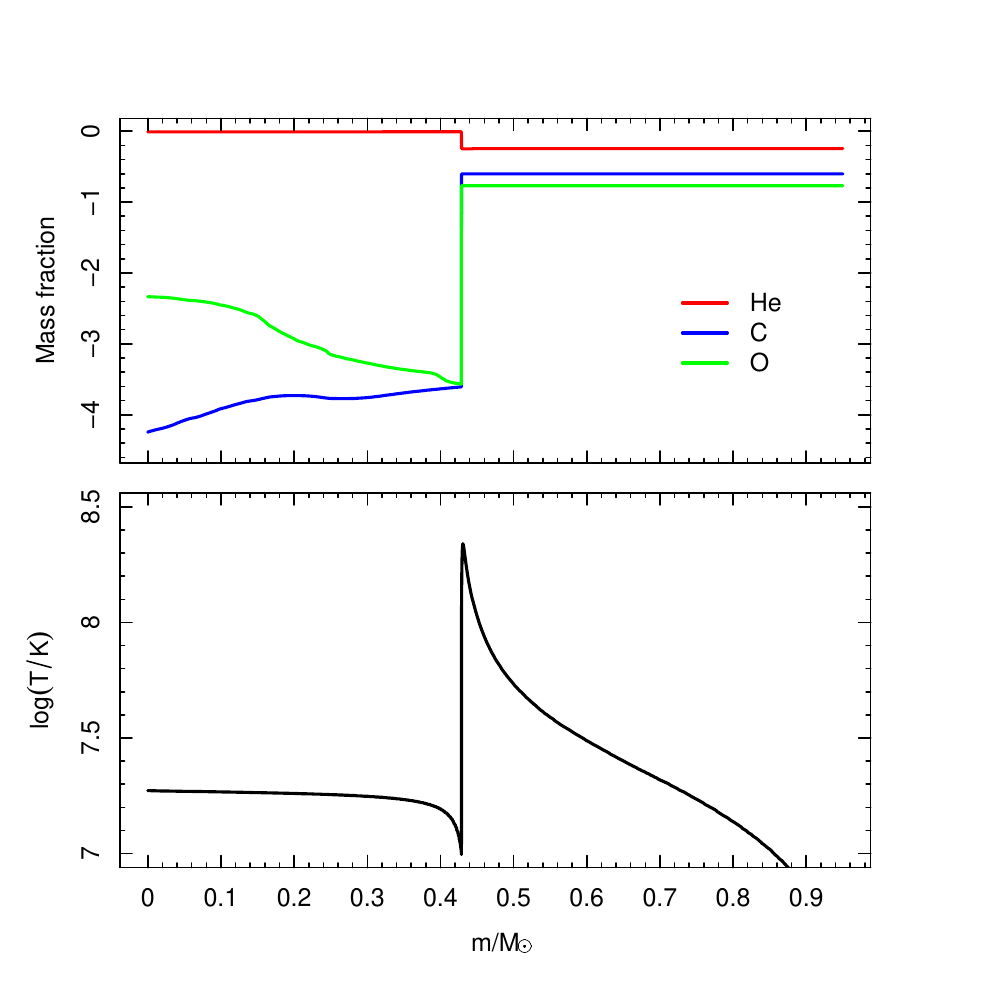}
    \caption{The chemical structure and the temperature profile of the post-merger model PG 1528+025 for 0.45$\Msun$ He WD and 0.50$\Msun$ sdB. The bottom panel is the temperature profile, and the top is the chemical structure.}
    \label{fig:12}
\end{figure}

For this reason, instead of attempting to predict the post-merger surface abundances directly, we adopt the observed abundances of the three CO-sdO stars as the composition of the accreted material, as shown Fig.~\ref{fig:12}. We then compute the accretion process and the subsequent stellar evolution to test whether the resulting evolutionary tracks are consistent with the observed stellar parameters ($T _ {\rm eff}, \rm log (g)$, and luminosity). In this sense, the calculations presented here should be regarded as a consistency test of the merger scenario rather than a fully predictive model of the surface abundances.

The resulting evolutionary tracks are shown in Figs.~\ref{fig:13}--\ref{fig:15}. In both the Kiel and HR diagrams, the theoretical tracks for several merger models pass close to the observed positions of the three CO-sdO stars. PG 1654+322 appears consistent with a stable core helium-burning phase, while PG 1528+025 and UCAC4 108 are located at somewhat higher luminosities, suggesting that they may already have entered the helium shell-burning phase.

\subsubsection{Effects of atomic diffusion}
To assess the potential influence of atomic diffusion on the predicted surface abundances, we further analyzed the He WD+sdB merger model for PG 1528+025. In this analysis, the initial surface abundances of C and O were set to the observed values of PG 1528+025, and their subsequent evolution was followed with diffusion included. The result is shown in Fig.~\ref{fig:16}. By the time central He burning begins, at \(t\simeq 9\) Myr, the surface C abundance decreases from about \(0.24\) to \(0.19\), while the surface O abundance decreases from about \(0.16\) to \(0.10\). Thus, diffusion reduces the C and O mass fractions by about \(0.05\)--\(0.06\), but does not completely erase the C- and O-rich surface composition before the onset of central He burning. The resulting abundances remain broadly consistent with the observed values of PG~1528+025 within the observational uncertainties.

After the onset of central He burning, the model enters a much longer evolutionary phase during which atomic diffusion continues. If no competing process is present, gravitational settling may further reduce the photospheric C and O abundances and may eventually lead to a nearly pure-He surface. This effect is important for interpreting the observed abundances of C- and O-rich hot subdwarfs. We would like to note that this calculation does not include hydrogen. The present model assumes that the residual hydrogen has been consumed or removed during the merger process. If a small amount of hydrogen survives the merger, gravitational settling could cause hydrogen to float to the surface more rapidly than C and O sink, potentially producing a H-rich surface layer on a shorter timescale. The predicted photospheric composition is therefore sensitive to the residual hydrogen mass, the efficiency of diffusion, and possible competing processes such as radiative levitation, turbulent mixing, and weak stellar winds. A fully self-consistent treatment of these processes is beyond the scope of the present work, but the present test shows that diffusion alone does not immediately destroy the C- and O-rich surface abundance pattern during the early post-merger evolution of PG~1528+025.

\subsection{Combined comparison with observations}

A comparison of the two merger scenarios presented above reveals several common features. Both the double sdB and He WD + sdB merger models produce evolutionary tracks that broadly match the observed positions of the three CO-sdO stars. In both the Kiel and HR diagrams, the theoretical tracks pass close to the observed stellar parameters within the observational uncertainties. This suggests that the carbon- and oxygen-rich surface compositions are most naturally explained by nucleosynthesis in the cores of sdB stars prior to the merger. Efficient mixing of the processed material into the outer layers is likely to occur during the merger. In both scenarios, the theoretical tracks indicate that the masses of the merger remnants are likely to be about 0.8–1.0$\Msun$. This is significantly higher than the canonical mass of typical sdB stars ($\approx 0.47 \Msun$).

\begin{figure}
    \centering
    \includegraphics[width=1\linewidth]{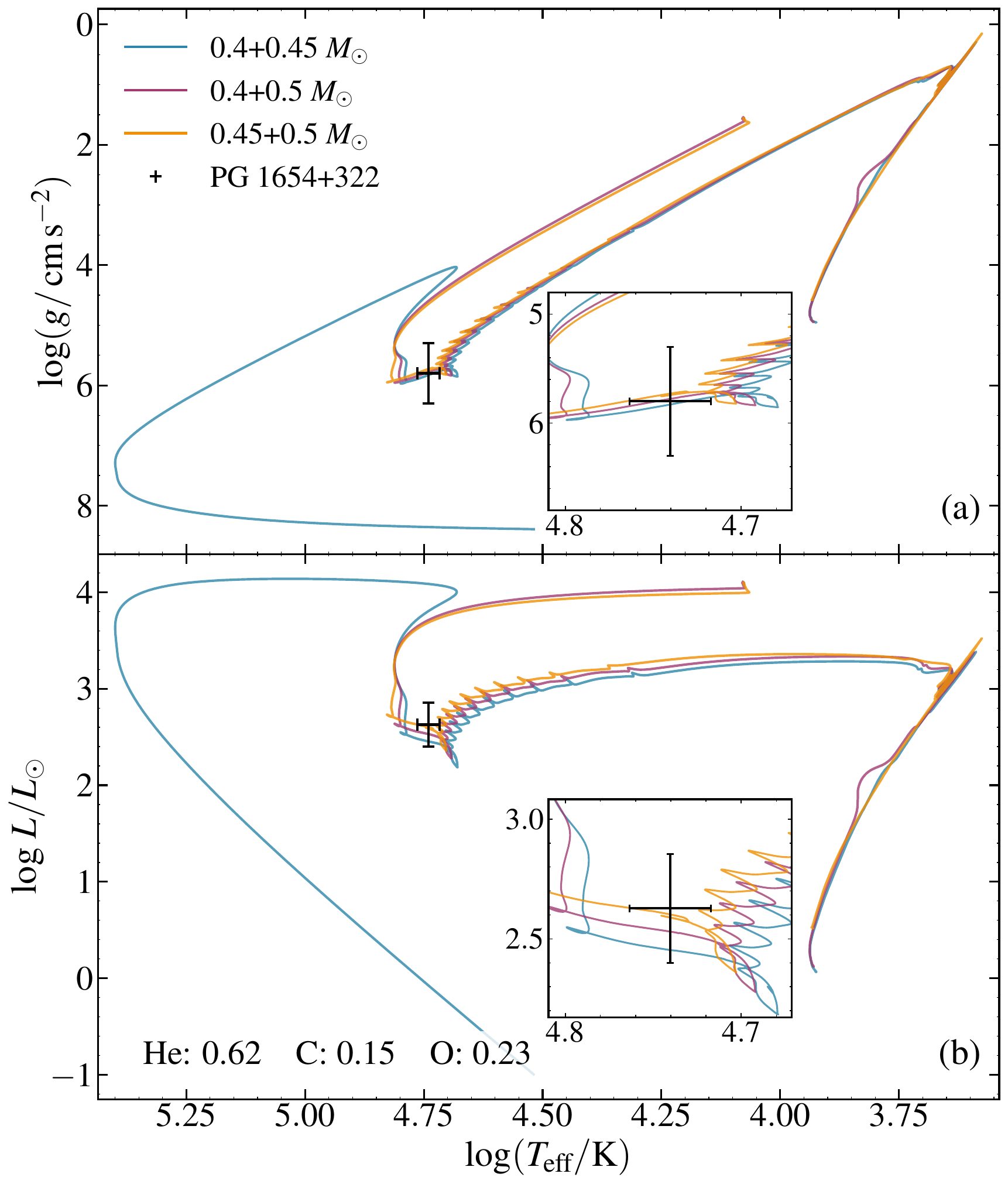}
    \caption{Evolutionary tracks for PG 1654+322 merger remnants in the sdB + He WD merger scenario. Panel (a) shows the Kiel diagram, and panel (b) shows the HR diagram. Black error bars indicate the observed ranges of the stellar parameters.}
    \label{fig:13}
\end{figure}

\begin{figure}
    \centering
    \includegraphics[width=1\linewidth]{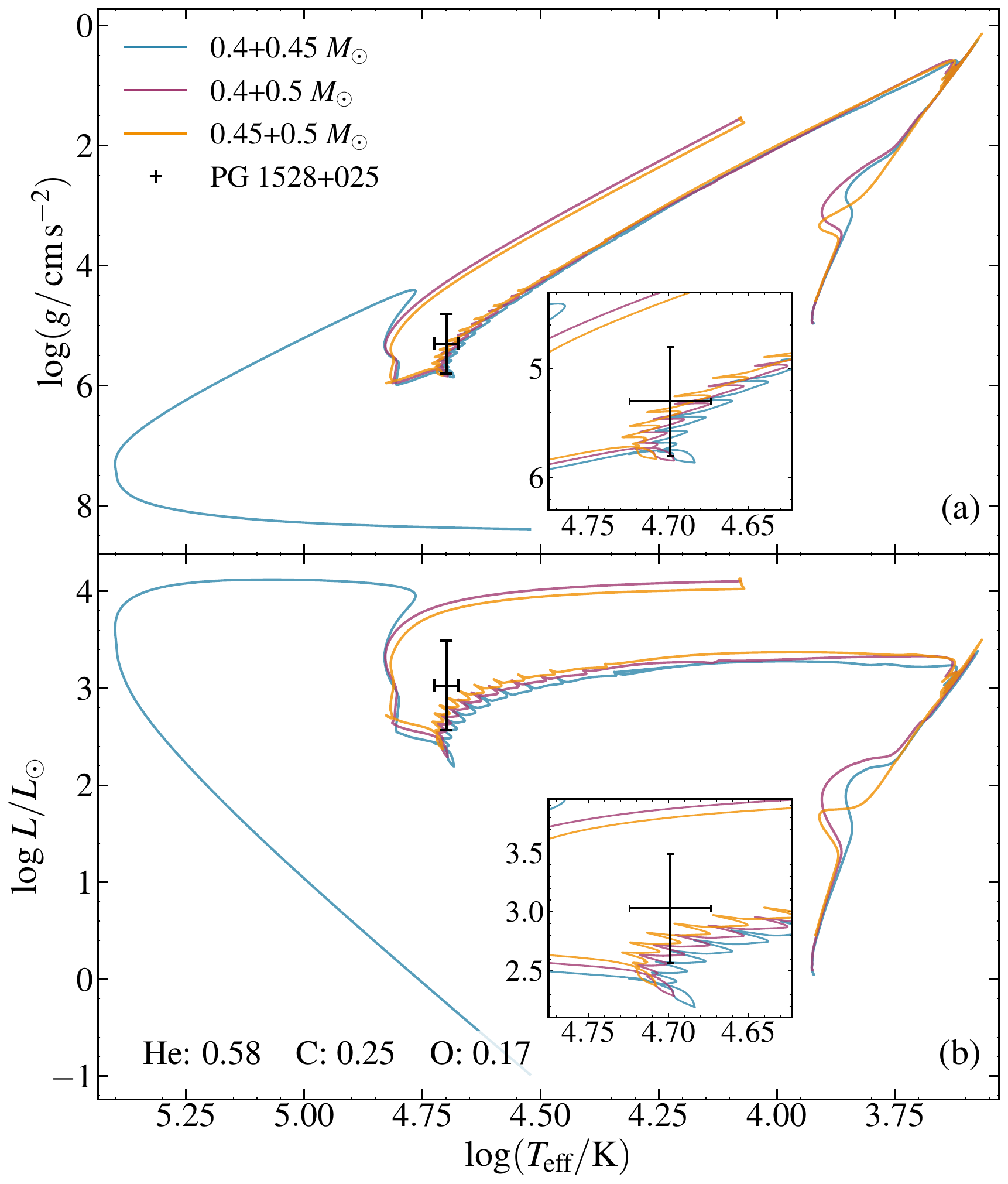}
    \caption{Same as Fig.~\ref{fig:13}, but for PG 1528+025.}
    \label{fig:14}
\end{figure}

\begin{figure}
    \centering
    \includegraphics[width=1\linewidth]{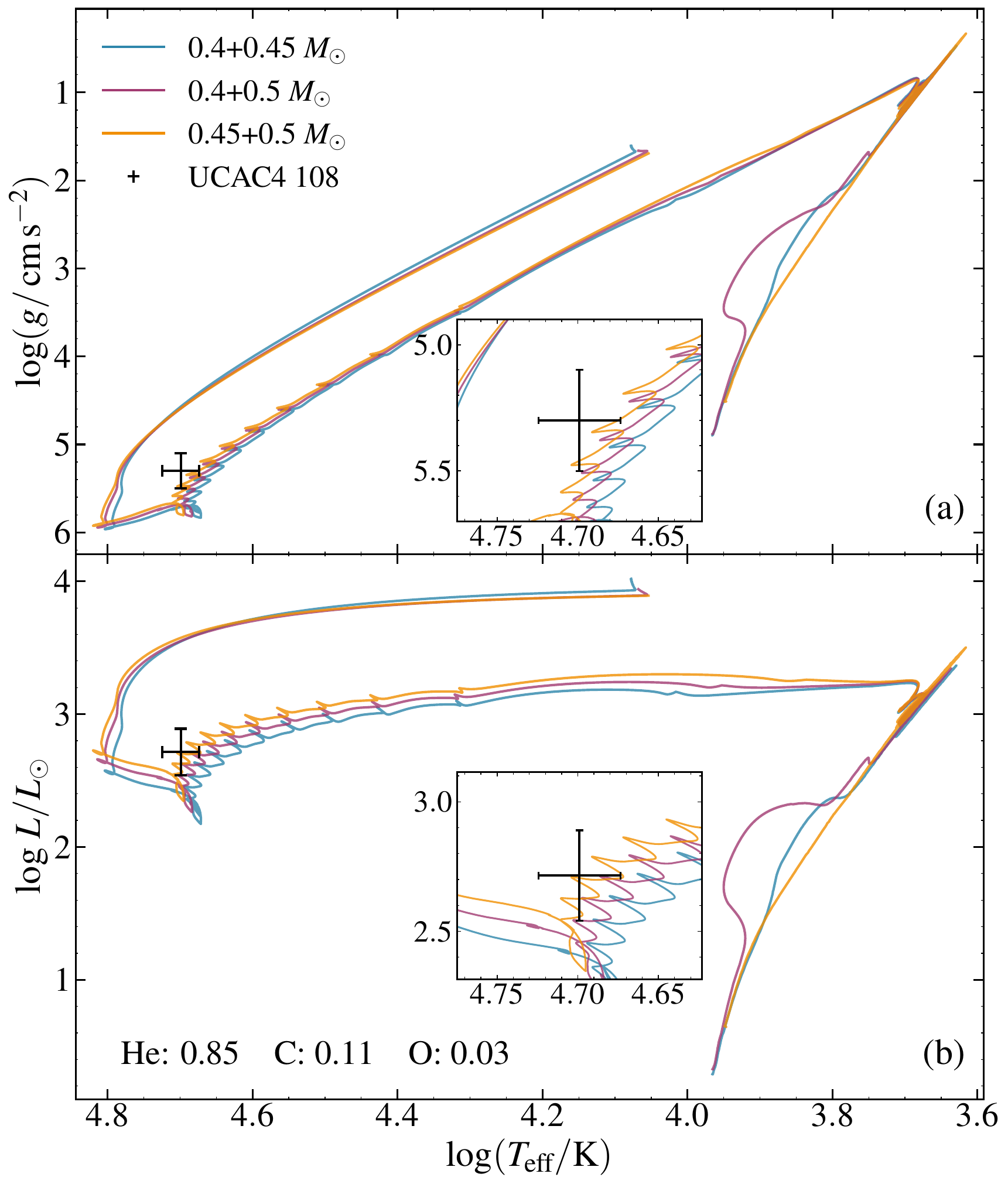}
    \caption{Same as Fig.~\ref{fig:13}, but for UCAC4 108.}
    \label{fig:15}
\end{figure}

\begin{figure}
    \includegraphics[width=\columnwidth]{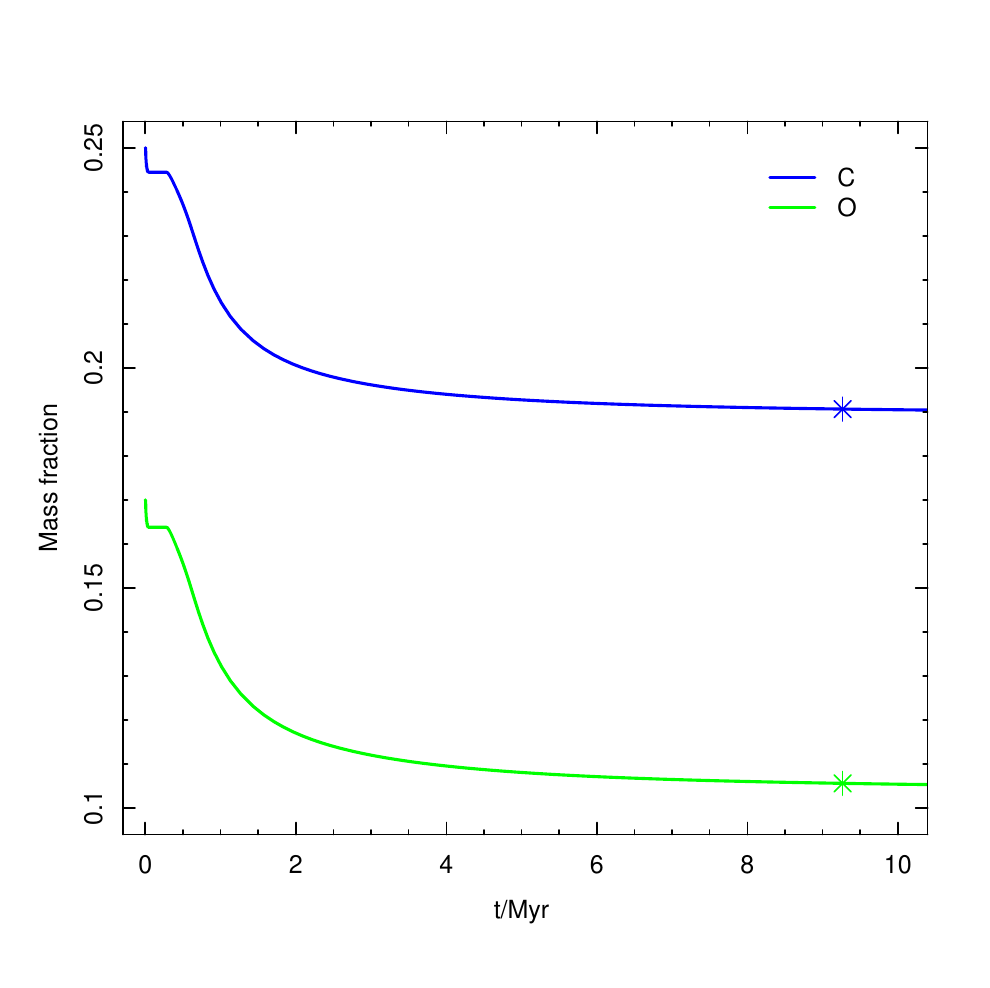}
    \caption{Evolution of the surface mass fractions of C and O for the He WD+sdB merger model corresponding to PG~1528+025, with atomic diffusion included. The initial surface C and O abundances are set to the observed values of PG~1528+025. The blue and green curves show the surface mass fractions of C and O, respectively. The asterisk marks the onset of central He burning.}
    \label{fig:16}
\end{figure}

Despite these similarities, the evolutionary tracks also provide clues about the evolutionary stages of the observed CO-sdO stars. PG 1654+322 appears consistent with a stable core helium-burning phase in both scenarios. In contrast, PG 1528+025 and UCAC4 108 are located at somewhat higher luminosities than those predicted for central helium burning. This interpretation is consistent with the expectation that shell-burning objects are more luminous than stars undergoing stable central helium burning.

Both merger channels are broadly consistent with the current observational constraints, but they differ in post-merger evolutionary timescales. In the double sdB scenario, the merger remnant can reach a stable core helium-burning configuration relatively quickly. This behaviour is similar to that expected in the formation of blue straggler stars through stellar mergers. By contrast, in the He WD + sdB scenario, the remnant undergoes a sequence of inward-propagating helium flashes before stable central helium burning is established. This process may take of order $10^7$ yr before the star settles into stable helium burning. Although diffusion has been included in our post-merger models, the short timescale between the merger and the onset of stable helium burning prevents significant elemental settling. This further suggests that adopting the observed abundances, or values slightly higher than them, as the initial conditions for the post-merger evolutionary calculations is consistent with the available observations.

An additional observational clue could help distinguish between these two scenarios. PG 1654+322 has been reported to show an infrared excess\citep{Werner2022}. This may indicate the presence of circumstellar material associated with a recent merger event. If this excess indeed originates from merger-related material, the event likely occurred relatively recently in evolutionary terms. In that case, the double sdB scenario may better explain the origin of this object. In this scenario, the remnant can reach the observed state shortly after the merger. However, current evidence is limited. Further infrared observations of CO-sdO stars will be required to determine whether such features are common among merger remnants and to better constrain the relative importance of different formation channels.

\subsection{Effects of global C/O enrichment on helium-star evolutionary tracks}

To examine how \(\mathrm{C/O}\) enrichment affects the evolutionary tracks, we performed a set of controlled calculations using chemically homogeneous helium-star models. In these tests, the chemical composition was modified throughout the entire helium-star model, rather than only in a surface layer or envelope. We separately enhanced the carbon and oxygen abundances and adopted the observed composition of PG\,1654+322. The models were then evolved self-consistently from central helium burning to the white-dwarf cooling phase.

Fig.~\ref{fig:17} shows the resulting evolutionary tracks in the HR diagram. Compared with the reference helium-star model, the C/O-enhanced models show noticeable shifts in effective temperature and luminosity. The model with the PG\,1654+322-like composition provides a useful comparison with the observed position of PG\,1654+322, although these calculations should be regarded as controlled tests rather than detailed merger-remnant models.

Replacing helium by carbon or oxygen changes the mean molecular weight, opacity, and internal structure of the star. The stellar model responds by adjusting its central temperature, density, and convective-core properties, so the resulting luminosity cannot be interpreted as a direct consequence of the change in helium abundance alone. Because helium burning is strongly temperature sensitive, the nuclear energy generation adjusts self-consistently with the stellar structure. The reduced helium abundance primarily affects the available fuel and the evolutionary timescale. The shifts of the tracks in Fig.~17 therefore reflect the combined effects of composition, opacity, mean molecular weight, internal structure, and convective-core evolution.

\begin{figure}
    \centering
    \includegraphics[width=1\linewidth]{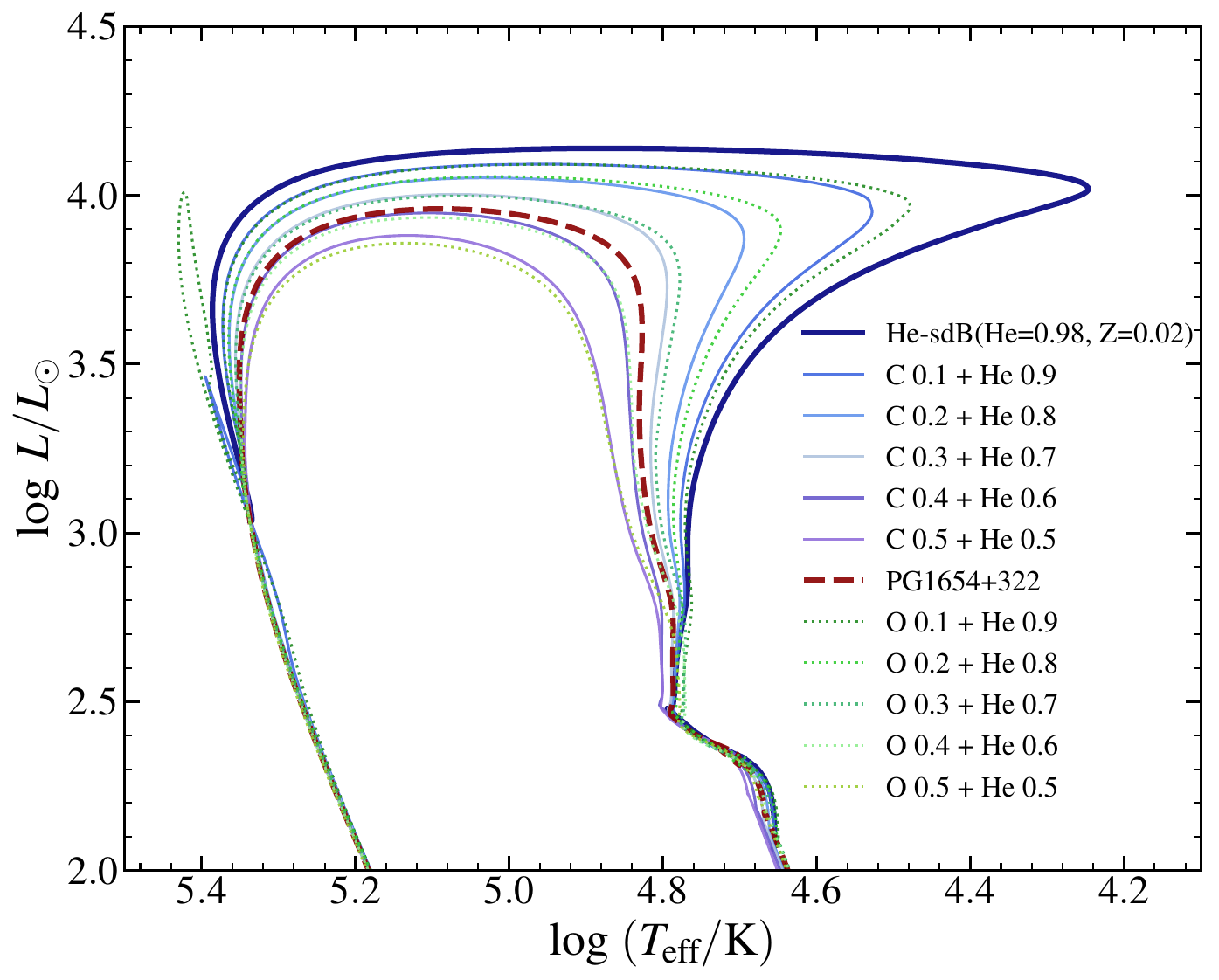}
    \caption{Evolutionary tracks of $0.8\,M_\odot$ chemically homogeneous models with different compositions in the Hertzsprung--Russell diagram. The reference He-sdB model is hydrogen-free and has $X_{\rm He}=0.98$ and $Z=0.02$. For the C-enhanced sequence, the labels give the carbon and helium mass fractions; for example, ``C 0.5 + He 0.5'' denotes $X_{\rm C}=0.5$ and $X_{\rm He}=0.5$. The O-enhanced sequence is defined analogously; for example, ``O 0.5 + He 0.5'' denotes $X_{\rm O}=0.5$ and $X_{\rm He}=0.5$.}
    \label{fig:17}
\end{figure}

\section{Conclusions and outlook}
CO-sdO stars represent a recently identified subtype of helium-rich hot subdwarfs exhibiting extreme enrichment in carbon and oxygen, whose evolutionary origin remains uncertain. In this work, we propose that the characteristic surface abundances of CO-sdO stars may arise from the mixing of carbon and oxygen synthesized in the cores of sdB stars during merger events, via either the double sdB or the He WD + sdB merger channel.
For these two formation channels, we draw the following conclusions:\\
1. For double sdB mergers, mixing sdB stars of different masses and evolutionary stages can yield carbon mass fractions as high as 0.25 and oxygen mass fractions up to 0.22. The corresponding evolutionary tracks agree well with the three currently known CO-sdO stars, although PG 1528+025 and UCAC4 108 exhibit luminosities slightly higher than predicted by the theoretical tracks.\\
2. The evolutionary tracks for the sdB + He WD merger channel are also broadly consistent with the observed luminosities, effective temperatures, and surface gravities of the three CO-sdO stars. In particular, PG 1528+025 and UCAC4 108 are more consistent with evolutionary stages corresponding to inward helium-shell burning.\\

An important caveat is that the double-sdB merger channel is expected to be intrinsically rare. Its formation requires two nearly coeval stars to develop non-degenerate helium cores, eject a common envelope, and emerge in an orbit compact enough for Roche-lobe contact during the core-helium-burning phase. The order-of-magnitude estimate of \citet{Justham2011} suggests that candidate double-core hot-subdwarf systems occur at roughly one per $3\times10^{3}$ normal sdB stars, and only a subset of these systems are expected to merge. We therefore do not attempt to assess the contribution of this channel to the overall CO-sdO population. These calculations demonstrate its physical feasibility for suitable progenitor systems; determining its formation probability and birth rate requires dedicated binary population-synthesis calculations.

Although our results demonstrate that sdB mergers can reproduce many of the observed properties of CO-sdO stars, several uncertainties remain. 

First, the merger process is modelled using a simplified one-dimensional treatment in MESA. In reality, stellar mergers are inherently three-dimensional hydrodynamic events that involve complex processes such as differential rotation, angular momentum transport, and incomplete mixing. Hydrodynamic simulations using smoothed particle hydrodynamics (SPH) could provide a more realistic description of the merger process and help verify the mixing efficiency assumed in this work.

Second, in our calculations, the accretion material of the sdB is assumed to be fully mixed. This treatment should be regarded as a first-order approximation rather than a realistic description of the merger process. The hydrodynamic phase of stellar mergers is intrinsically three-dimensional. The extent of mixing cannot be reliably determined in one-dimensional stellar evolution calculations. The fully mixed configuration adopted here represents a simplifying assumption. It allows us to explore the subsequent thermal evolution of the remnant.

Physically, the assumption of complete mixing may be viewed as an upper limit to the efficiency with which the carbon- and oxygen-rich material produced in the helium-burning core of the sdB star can be transported to the outer layers. If mixing during the merger were less efficient, the surface abundances of carbon and oxygen would likely be lower than in the fully mixed case. In this sense, the present calculations mainly test whether the interior of a sdB star can provide a sufficiently large reservoir of CO-rich material. This helps to account for the observed surface compositions of the CO-sdO stars.

Consequently, our results should not be interpreted as implying that real merger remnants are necessarily fully mixed. Instead, the agreement between some of our models and the observed properties of the three stars indicates a specific feature. If a substantial fraction of the sdB core material is mixed outward during the merger, the resulting remnants can naturally reproduce both the chemical compositions and the positions of these objects in the Kiel and HR diagrams.

Finally, uncertainties in key stellar-evolution parameters, such as convective-mixing efficiency, overshooting, and diffusion, may affect the predicted abundances and evolutionary tracks. A systematic exploration of the parameter space would help quantify these uncertainties.

Overall, our calculations show that merger remnants involving sdB stars provide a plausible evolutionary pathway for the formation of CO-rich hot subdwarfs. In particular, the helium-burning core of the sdB star can act as a natural reservoir of carbon- and oxygen-rich material, which may be mixed into the outer layers during the merger and subsequently give rise to the observed surface composition of CO-sdO stars. Both the double sdB and He WD +sdB channels reproduce, at least for part of the evolutionary sequence, the observed positions of the three known CO-sdO stars in the Kiel and HR diagrams. However, we emphasize that the present calculations should be regarded as a test of physical consistency rather than as a unique identification of the formation pathway for any individual object. In particular, the assumed degree of mixing during the merger remains uncertain, and a more realistic treatment of the hydrodynamic merger process will be required to determine how efficiently the CO-rich core material can be transported to the surface.

If CO-sdOs are indeed produced through mergers involving sdB stars, several observational signatures may help to test this scenario. The merger remnants are expected to have higher masses than typical sdB stars, potentially reaching $\sim 0.8-1.0 \Msun$, and they may also retain relatively rapid rotation or strong magnetic fields as a consequence of angular-momentum redistribution during the merger. More importantly, this evolutionary channel is closely linked to common-envelope evolution, since the formation of close sdB binaries capable of merging requires prior envelope ejection. The birthrate of CO-sdOs should therefore depend sensitively on the population of such close sdB binaries. A detailed assessment of this channel will require binary population-synthesis calculations, which we plan to present in future work to derive more quantitative estimates of the expected formation probability and birth rate of CO-sdO stars. At present, the population studies of hot subdwarf binaries by \cite{Han2002,Han2003} and \cite{Justham2011}  provide useful preliminary guidance on the possible frequency of these evolutionary pathways.

\section*{Acknowledgements}

We express gratitude to the anonymous referee, whose comments helped improve the quality of this work.This work is supported by grant Nos. 12473028 and 12541303 from the National Natural Science Foundation of China, and Beijing Undergraduate Students Innovation and Entrepreneurship Training Program of China (No.X202510027292).

\section*{Data Availability}

The MESA inlists, initial stellar models, and main evolutionary outputs used in this work are available on Zenodo.



\bibliographystyle{mnras}
\bibliography{mybib} 



\bsp	
\label{lastpage}
\end{document}